\documentclass[aps,amssymb,floatfix,prd,amsmath,preprintnumbers]{revtex4}
\usepackage{graphicx}  
\usepackage[font=scriptsize]{caption}
\usepackage{float}
\usepackage{dcolumn}   
\usepackage{bm}
\usepackage{wasysym}
\usepackage[T1]{fontenc}
\usepackage{color}
\begin{document}

\title{\bf  Nuclear Symmetry Energy and Neutron Skin Thickness of $^{208}$Pb using a finite range effective interaction }

\author{ D. Behera\footnote{1. Department of Physics, Indira Gandhi Institute of Technology, Sarang, Dhenkanal, Odisha-759146, India\\2. School of Physics, Sambalpur University, Jyotivihar, Sambalpur, Odisha-768019, India,\\E-mail:dipadolly@rediffmail.com}, S. K. Tripathy\footnote{Department of Physics, Indira Gandhi Institute of Technology, Sarang, Dhenkanal, Odisha-759146, India, E-mail:tripathy\_ sunil@rediffmail.com}, T. R. Routray \footnote{Retired Professor, School of Physics, Sambalpur University, Jyotivihar, Sambalpur, Odisha-768019, India, E-mail:trr1@rediffmail.com}and B. Behera\footnote{Retired Professor, School of Physics, Sambalpur University, Jyotivihar, Sambalpur, Odisha-768019, India} 
}
\affiliation{ }

\begin{abstract}
We use a finite range simple effective interaction to construct nuclear equations of state for the study of the density dependence of the nuclear symmetry energy. The EoSs provide good descriptions of the nuclear symmetry energy at a subsaturation density $\rho_c=0.11$ fm$^{-3}$ and at a density around two times the saturation density $\rho_0$.  We obtain a correlation between the neutron skin thickness in $^{208}$Pb and the density slope parameter at the subsaturation density. A linear relation is obtained between the neutron skin thickness and the parameter $\beta^{\prime}=\frac{L(\rho_c)}{3E_s(\rho_0)}$, where $E_s(\rho_0)$ and $L(\rho_c)$ are respectively the nuclear symmetry energy at saturation density and the density slope parameter at the subsaturation density. 
\end{abstract}
\maketitle
\section{Introduction} 

The nuclear symmetry energy (NSE) $E_s(\rho)$, defined as the energy cost per particle for changing all the protons into neutrons in symmetric nuclear matter (SNM), is a fundamental quantity in nuclear physics and astrophysics. It provides a link to the properties of atomic nuclei with the structure and dynamics of neutron stars. An accurate knowledge of NSE is essential to understand the equation of state of (EoS) of isospin asymmetric nuclear matter (ANM). In fact, the understanding and constraining the EoS of dense neutron-rich matter has remained a major goal in nuclear science research \cite{Li2019a}. In neutron stars, NSE controls the proton fraction of the beta stable matter and decides the cooling mechanism and thickness of the neutron star crusts \cite{Lattimer2000, Steiner2005, Steiner2012}. Also, $E_s(\rho)$ has a great role in obtaining the mass-radius relation and tidal deformability in neutron stars \cite{Ji2019}. In astrophysics, its behaviour is required to understand the supernova explosion mechanism  and stellar nucleosynthesis \cite{Lattimer2007, Klahn2006, Loan2011}. In nuclear physics, NSE  has important consequences for understanding the dynamics in heavy-ion reaction processes involving isospin asymmetric nuclei, prediction of properties of exotic nuclei with large neutron excess, etc. Experimentally, $E_s(\rho)$ is not a directly measurable quantity and has to be extracted from isospin sensitive observables. The experimental determination of NSE therefore depends on the reliability of the model in describing the experimental observables. Although, it remains a quite challenging task, in the last one and half decade, a lot of theoretical and experimental efforts have been made to constrain the density dependence of NSE.  In recent years, significant progress has been made to constrain the NSE at low densities from dynamical behaviour \cite{Tsang2009, Tsang2012}, resonances and excitations \cite{Piek2012, Tamii2011, Maza2013, Zhang2014}, static properties of finite nuclei \cite{Gor2009, Wang2014, Tian2014, Chen2015},  neutron skin thickness \cite{Warda2009, Vinas2014, Mondal2016, Wang2013, Zhang2013, BKA12, BKA13} and electric dipole polarisability \cite{Zhang2014, Hashimoto2015,Tonchev2017, Tamii2014}. The density dependence of $E_s(\rho)$ is largely unknown, except the value of $E_s(\rho_0)$  and its slope parameter $L(\rho_0)$ at saturation density $\rho_0$. While data from nuclear experiments and astrophysical observations prior to 2013 constrain these two parameters as $E_s(\rho_0)=31.6\pm 2.7$ MeV and $L(\rho_0)=58.9\pm 16$ MeV \cite{Li2013}, some recent work extracted the values  $E_s(\rho_0)=31.7\pm 3.2$ MeV and $L(\rho_0)=58.7\pm 28.1$ MeV \cite{Oertel2017}. However, our present knowledge of the curvature parameter $K_{sym}(\rho_0)$ is rather poor. Usually, one needs to assume some models and then the data of stable nuclei to the parameters of the assumed models are fitted. This process may not constrain the isovector part of the nuclear equations of state in a precise way \cite{Vinas2014}.

Neutron skin thickness (NST) has proved to be a better tool to improve our knowledge in the isovector channels of nuclear effective interaction \cite{Trzcinska2001, Brown2007, Klos2007,Zenihiro2010, Roca-Maza2012}. The Lead Radius Experiment (PREX) is able to determine the neutron radius in $^{208}$Pb up to $1\%$ accuracy through the measurement of the parity violating asymmetry at low momentum transfer in the polarised elastic electron scattering performed at an angle $5^0$\cite{prex2012}. This experiment provides the first purely electroweak, model-independent measurement of the weak charge form factor which is closely related to the neutron skin thickness of $^{208}$Pb. A measurement of the form factor of $^{208}$Pb at a momentum transfer of $q\simeq 0.475$ fm$^{-1}$, yielded the PREX results $\varDelta r_{np}=0.33^{+0.16}_{-0.18}$ fm \cite{prex2012}. Another run of the same has been proposed (PREX-II) to reduce the uncertainty to $0.06$ fm \cite{prex2}. The nuclear droplet model (DM) \cite{Myers1980} suggested that NST in heavy nucleus is related to different symmetry energy parameters. On the basis of DM, it has been predicted from different mean field models in $^{208}$Pb that, NST is linearly correlated with the density slope parameter \cite{Maza2011}. In recent years, different correlation systematics of isovector properties of finite nuclei with NST have yielded constraints on the density slope parameter and the curvature symmetry parameter \cite{Warda2009, Vinas2014, Maza2011, Mondal2016, Centelles2010}. However so far a consensus has not yet been reached in this direction.

The purpose of the present work is two fold. First, we construct nuclear equations of state using a finite range effective interaction so that the EoSs reproduce the basic features of SNM at saturation density and the nuclear symmetry energy at the subsaturation cross density $\rho_c=0.11$ fm$^{-3}$ as constrained in a recent work from analysing the binding energy difference of heavy isotope pairs\cite{Zhang2013}. Similar method has been adopted in a recent work \cite{Zhou2019} to constrain the density dependence of NSE at high density. Second, through a comprehensive and quantitative analysis, we correlate the neutron skin thickness  and the energy constant of isovector giant dipole resonance (IVGDR) of  $^{208}$Pb to the isovector indicators of the finite range effective interaction.  The paper is organised as follows: in Section II, the basic formalism of the finite range effective interaction and the method of parameter fixation are discussed. In Section III, we discuss the mass dependence of symmetry energy coefficient of finite nuclei. In Section IV, we carry out a comprehensive analysis for the correlation of the neutron skin thickness with the density slope parameter at saturation density and at a reference density $\rho_c=0.11$ fm$^{-3}$. The correlation of IVGDR energy constant of $^{208}$Pb with the isovector indicators of the effective interaction is presented in Section-V. At the end, the conclusion and summary of the present work are presented in Section-VI.

\section{Model and Method}
\subsection{Nuclear Symmetry Energy and Nuclear Equation of state}
For an isospin asymmetric nuclear matter (ANM) with neutron-proton asymmetry $\delta$, neutron density $\rho_n$, proton density $\rho_p$, the energy per nucleon $e(\rho,\delta)$ can be written as \cite{Behera2016}
\begin{equation}
e(\rho,\delta)=\frac{1}{\pi^2\rho}\left[f(k_n)+f(k_p)\right]+V(\rho,\delta)\label{eq:1},
\end{equation}
where $\rho=\rho_n+\rho_p$ is the nucleon density, $k_{n,p}$ are the neutron(proton) Fermi momenta. $V(\rho,\delta)$is the interaction part of the equation of state, $e(\rho,\delta)$. The kinetic part of the EoS is treated in the relativistic Fermi gas model and the functional $f(k_i)$ with $i=n,p$ is expressed as \cite{Behera2016}
\begin{equation}
f(k_i)=\int\limits_{0}^{k_f}\left(c^2\hbar^2k^2+M^2c^4\right)^{\frac{1}{2}}k^2dk \label{eq:2}.
\end{equation}
Here, $Mc^2$ is the rest mass of the nucleon. The interaction part of $e(\rho,\delta)$ has a complicated dependence on $\rho$ and $\delta$ because of the presence of finite range exchange interaction between nucleons. However, the isospin exchange symmetry allows us to expand the EoS in even powers of the neutron-proton asymmetry $\delta$ as 
\begin{equation}
e(\rho,\delta)=e_0(\rho)+E_s(\rho)\delta^2+\mathcal{O}(\delta^4)\label{eq:3},
\end{equation}
where $e_0(\rho)=e(\rho,\delta=0)$ is the energy per particle in symmetric nuclear matter (SNM). $E_s(\rho)$ is nuclear symmetry energy defined as

\begin{equation}
E_s(\rho)=\frac{1}{2!}\frac{\partial^2e(\rho,\delta)}{\partial\delta^2}|_{\delta=0} \label{eq:4}.
\end{equation}
Assuming that the contribution from higher order terms in $\delta$ is small, the NSE can be expressed as the difference in the energy per particle in pure neutron matter $e_n(\rho)= e(\rho,\delta=1)$ and that in SNM,
\begin{equation}
E_s(\rho)=e_n(\rho)-e_0(\rho)\label{eq:5}.
\end{equation}
An expansion of  $E_s(\rho)$ around the saturation density $\rho_0$ reads as 
\begin{eqnarray}
E_s(\rho) &\approx& E_s(\rho_0)+\frac{L}{3}\left(\frac{\rho-\rho_0}{\rho_0}\right)
         +\frac{K_{sym}}{18}\left(\frac{\rho-\rho_0}{\rho_0}\right)^2+\cdots \label{eq:6},
\end{eqnarray}
where $L(\rho_0)=3\rho_0\frac{\partial E_s(\rho)}{\partial \rho}|_{\rho=\rho_0}$ and $K_{sym}(\rho_0)=9\rho_0^2\frac{\partial^2 E_s(\rho)}{\partial \rho^2}|_{\rho=\rho_0}$ are the slope and curvature parameters of $E_s(\rho)$ at $\rho_0$. Up to 2nd order in the deviation from saturation density, the density dependence of NSE depends on the parameters $L(\rho_0)$ and $K_{sym}(\rho_0)$. 
\subsection{Finite range effective interaction}
In the present work, we consider a finite range simple effective interaction (SEI)
\begin{equation}\label{eq:7}
v_{eff}(\textbf {r})=t_0(1+x_0P_{\sigma})\delta(\textbf {r})+\frac{1}{6}t_3(1+x_3P_{\sigma})\left[\frac{\rho(\textbf {R})}{1+b\rho(\textbf {R})}\right]^{\gamma}\delta (\textbf {r})+\left(W+BP_{\sigma}-HP_{\tau}-MP_{\sigma}P_{\tau}\right)f(r),
\end{equation}
where $f(r)$ represents the form factor of a finite range Yukawa interaction $\frac{e^{-r/{\alpha}}}{r/{\alpha}}$. Here $\alpha$ is the range of the interaction. $\textbf{r}$ and $\textbf{R}$ are respectively the relative and centre of mass coordinates of the two interacting nucleons. $W, B, H$ and $M$ are the strength of the Wigner, Bartlett, Heisenberg and Majorana components.  $P_{\sigma}$ and $P_{\tau}$ are the spin and isospin exchange operators respectively. The simple effective interaction is similar to the Skyrme type interaction except the fact that the $t_{1}$  and $t_2$ terms of the Skyrme interaction are replaced by a finite range term. Also, the $t_3$ term has been modified. The replacement of the $t_{1}$  and $t_2$ terms by a finite range term is essential to account for the correct momentum dependence of the nuclear mean field as extracted from the optical model fits in heavy-ion collision studies at intermediate energies \cite{Berstch1988,Kuper1974,Pan1993, Mota1992,Zhang1994, Haddad1995, Dan1998, Dan2000}. In the denominator of the $t_3$ term we have considered a factor $\left(1+b\rho(\textbf{R})\right)^{\gamma}$ to avoid the supraluminous behaviour of the nuclear matter equation of state at high density region. The simple effective interaction contains altogether 11 adjustable parameters namely $t_0, x_0, t_3, x_3, b, \gamma, W, B, H, M$ and $\alpha$. This SEI has already been used to study the momentum and density dependence of the isoscalar part of the nuclear mean field at zero and finite temperature \cite{Behera1998, Behera2002, Routray2000}, isovector part of the nuclear mean field at zero temperature \cite{Behera2005} and the temperature dependence of nuclear symmetry energy \cite{Behera2009, Behera2011}. The SEI has also been used to calculate the half-lives of spherical proton emitters \cite{Routray2011}. 

The energy density $H(\rho,y_p, T)$ in ANM at a density $\rho$, proton fraction $y_p$ and temperature $T$ can be obtained from SEI as
\begin{eqnarray}
H(\rho,y_p, T) &=& \int \left[f_T^n(\textbf{k})+f_T^p(\textbf{k})\right]\left(c^2\hbar^2k^2+M^2c^4\right)~d^3k\nonumber\\
               &+& \frac{1}{2}\left[\frac{\varepsilon_0^l}{\rho_0}+\frac{\varepsilon_{\gamma}^l}{\rho_0^{\gamma+1}}\left(\frac{\rho}{1+b\rho}\right)^{\gamma}\right]\left(\rho_n^2+\rho_p^2\right)+\left[\frac{\varepsilon_0^{ul}}{\rho_0}+\frac{\varepsilon_{\gamma}^{ul}}{\rho_0^{\gamma+1}}\left(\frac{\rho}{1+b\rho}\right)^{\gamma}\right]\rho_n\rho_p\nonumber\\
               &+& \frac{\varepsilon_{ex}^{l}}{2\rho_0} \int\int \left[f_T^n(\textbf{k})f_T^n(\textbf{k}^{\prime})+f_T^p(\textbf{k})f_T^p(\textbf{k}^{\prime})g_{ex}(|\bf{k}-\bf{k}^{\prime}|)\right]~d^3k~d^3k^{\prime}\nonumber\\
               &+& \frac{\varepsilon_{ex}^{ul}}{2\rho_0} \int\int \left[f_T^n(\textbf{k})f_T^p(\textbf{k}^{\prime})+f_T^p(\textbf{k})f_T^n(\textbf{k}^{\prime})g_{ex}(|\bf{k}-\bf{k}^{\prime}|)\right]~d^3k~d^3k^{\prime}\label{eq:8},
\end{eqnarray}
where $f_T^{\tau}(\textbf{k}), \tau=n,p$ are the respective Fermi-Dirac distribution functions and $g_{ex}(|{\bf{k}}-{\bf{k}}^{\prime}|)=\frac{1}{1+\frac{|{\bf{k}}-{\bf{k}}^{\prime}|^2}{\Lambda^2}}$. $\Lambda$ is inversely related to the range parameter i.e. $\Lambda=\frac{1}{\alpha}$. The study of ANM involves nine parameters: $\gamma, b, \varepsilon_{0}^{l}, \varepsilon_{0}^{ul}, \varepsilon_{\gamma}^{l}, \varepsilon_{\gamma}^{ul}, \varepsilon_{ex}^{l}, \varepsilon_{ex}^{ul}$ and the range parameter $\alpha$. The parameters $\varepsilon_{0}^{l}, \varepsilon_{0}^{ul}, \varepsilon_{\gamma}^{l}, \varepsilon_{\gamma}^{ul}, \varepsilon_{ex}^{l}, \varepsilon_{ex}^{ul}$ are related to the parameters of SEI \cite{Behera2007}. 

The energy per particle in symmetric nuclear matter at zero temperature ($T=0$) for the Yukawa type finite range effective interaction becomes
\begin{equation}
e_0(\rho)=\frac{3Mc^2}{8x_f^3}\left[2x_fu_f^3-x_fu_f-ln\left(x_f+u_f\right)\right]+\frac{\varepsilon_0}{2}\frac{\rho}{\rho_0}+\frac{\varepsilon_{\gamma}}{2}\frac{\rho}{\rho_0^{\gamma+1}}\left(\frac{\rho}{1+b\rho}\right)^{\gamma}+\frac{\varepsilon_{ex}}{2\rho_0}\rho J_0(\rho)\label{eq:9},
\end{equation}
where $x_f=\frac{\hbar k_f}{Mc}$, $u_f=\left(1+x_f\right)^{\frac{1}{2}}$ and $k_f=\left(1.5\pi^2\rho\right)^{\frac{1}{3}}$ is the Fermi momentum in SNM. The functional $J_0(\rho)$ is given by
\begin{equation}
J_0(\rho) = \frac{\int\left(\frac{3j_1(k_fr)}{k_fr}\right)^2\frac{e^{-r/{\alpha}}}{r/{\alpha}}d^3r}{\int \frac{e^{-r/{\alpha}}}{r/{\alpha}}d^3r}\label{eq:10}.
\end{equation}
Here  $j_1(k_fr)$ is the first order spherical Bessel function and $\varepsilon_0=\frac{1}{2}\left(\varepsilon_{0}^{l}+ \varepsilon_{0}^{ul}\right), \varepsilon_{\gamma}=\frac{1}{2}\left(\varepsilon_{\gamma}^{l}+ \varepsilon_{\gamma}^{ul}\right)$, $\varepsilon_{ex}=\frac{1}{2}\left(\varepsilon_{ex}^{l}+ \varepsilon_{ex}^{ul}\right)$.

The energy per particle in pure neutron matter (PNM) at zero temperature is obtained from the SEI as
\begin{equation}
e_n(\rho)=\frac{3Mc^2}{8x_n^3}\left[2x_nu_n^3-x_nu_f-ln\left(x_n+u_n\right)\right]+\frac{\varepsilon_0^l}{2}\frac{\rho}{\rho_0}+\frac{\varepsilon_{\gamma}^l}{2}\frac{\rho}{\rho_0^{\gamma+1}}\left(\frac{\rho}{1+b\rho}\right)^{\gamma}+\frac{\varepsilon_{ex}^l}{2\rho_0}\rho J_{n}(\rho)\label{eq:11},
\end{equation}
where $x_n=\frac{\hbar k_n}{Mc}$, $u_n=\left(1+x_n\right)^{\frac{1}{2}}$ and $k_n=\left(3\pi^2\rho\right)^{\frac{1}{3}}$ is the Fermi momentum in PNM. The functional $J_{n}(\rho)$ is given by
\begin{equation}
J_{n}(\rho) = \frac{\int\left(\frac{3j_1(k_nr)}{k_nr}\right)^2\frac{e^{-r/{\alpha}}}{r/{\alpha}}d^3r}{\int \frac{e^{-r/{\alpha}}}{r/{\alpha}}d^3r}\label{eq:12}.
\end{equation}

\subsection{Constraining the interaction parameters}
The complete description of SNM requires only the knowledge of six parameters $\gamma, b, \alpha, \varepsilon_0, \varepsilon_{\gamma}$ and $\varepsilon_{ex}$. However, the equation of state in PNM requires the splitting of the strength parameters $\varepsilon_0, \varepsilon_{\gamma}$ and $\varepsilon_{ex}$ into like ($l$) and unlike ($ul$) channels. The parameters $\alpha$ and $\varepsilon_{ex}$ are obtained from a simultaneous optimization procedure so as to provide a correct momentum and density dependence of the nuclear mean field in SNM as demanded by the optical model fits to the heavy-ion collision data at intermediate energies. During the optimization procedure, it is kept in view that, the nuclear mean field in SNM at saturation density vanishes for a kinetic energy of $300~\text{MeV}/n$. $\varepsilon_{0}$ and $\varepsilon_{\gamma}$ are determined from the saturation condition in normal nuclear matter. Here we have used the fact that, $Mc^2=939$ MeV, energy per nucleon in SNM $e_0(\rho_0)=923$ MeV, $\left(c^2\hbar^2k_{f_0}^2+M^2c^4\right)^{\frac{1}{2}}=976$ MeV and the saturation density $\rho_0=0.1658$ fm$^{-3}$. The stiffness of the equation of state in SNM determines the exponent $\gamma$. In the present work, we have used $\gamma=\frac{1}{2}$ corresponding to the incompressibility in normal nuclear matter $K=240$ MeV. The parameters as constrained from different physical basis to determine the EoS of SNM are given in Table I.

\begin{table}
\caption{Values of the interaction parameters in SNM.}
\centering
\begin{tabular}{c|c|c|c|c|c|c}
\hline
\hline
Parameters & $\gamma$& $b$        & $\alpha$    & $\varepsilon_{0}$ & $\varepsilon_{\gamma}$ & $\varepsilon_{ex}$\\
in SNM	   &	     & (fm$^{3})$ & (fm)      & (MeV)	        & (MeV)                & (MeV)           \\
\hline
           &$0.5$&0.5668&0.4044&-57.86&76.91&-121.84\\

\hline
\end{tabular}
\end{table}

\begin{table}
\caption{Values of the interaction parameters in PNM corresponding to three different values of $E_s(\rho_0)$ and $E_s(\rho_c)=26.65$ MeV. The table includes the predicted values of $E_s^{\prime}(\rho_0)$ and $E_s(2\rho_0)$. }
\centering
\begin{tabular}{c|c|c|c|c|c|c}
\hline
\hline
Parameter sets & $E_s(\rho_0)$&    $E_s^{\prime}(\rho_0)$ & $E_s(2\rho_0)$ & $\varepsilon_{0}^{l}$ & $\varepsilon_{\gamma}^{l}$ & $\varepsilon_{ex}^{l}$\\
in PNM	   & (MeV)	      & (MeV)               & (MeV)& (MeV)	        & (MeV)                & (MeV)           \\
\hline
Set I      & 33               &   14.72    & 38.65  &  - 1.7401       &  21.0099 & -81.2299\\

Set II	   & 34               &   18.35    & 44.76  &  -11.2489      &  33.0473 & -81.2299\\
Set III    & 35               &   22       & 50.90  &  -20.8451      &  45.1762 &-81.2299\\

\hline
\end{tabular}
\end{table}

\begin{figure}[t]
\includegraphics[width=0.7\textwidth]{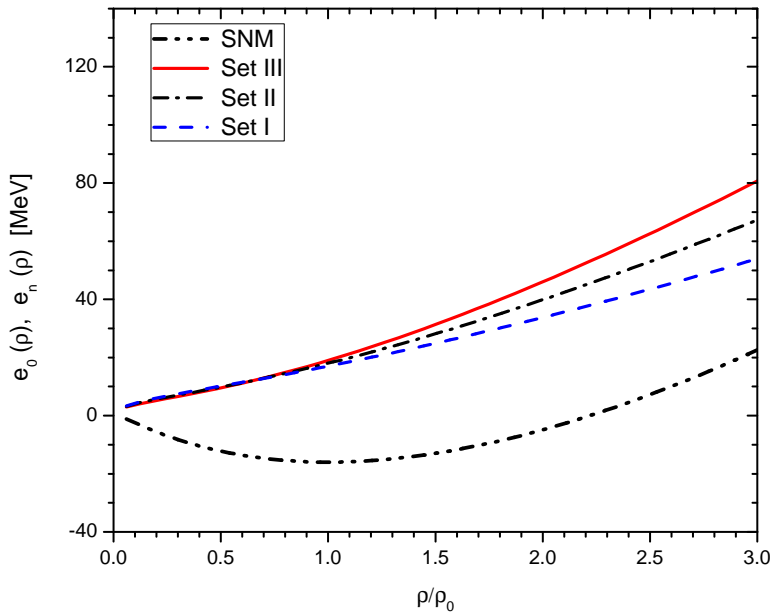}
\caption{Energy per particle in SNM and PNM for three different parameter sets of the finite range simple effective interaction. The dotted curve is for SNM and the solid curves represent the energy per particle in PNM.}
\end{figure}

There are no experimental or empirical constraints on the splitting of $\varepsilon_0, \varepsilon_{\gamma}$ and $\varepsilon_{ex}$ into the corresponding like ($l$) and unlike ($ul$) components. However, from an analysis of the entropy density in PNM and SNM, $\varepsilon_{ex}^l$ has been constrained in a recent work as $\varepsilon_{ex}^l=\frac{2}{3}\varepsilon_{ex}$ \cite{Behera2009}. For this constrained value of $\varepsilon_{ex}^l$, the neutron effective mass in neutron-rich matter is predicted to pass over the proton effective mass. Even though controversy prevails among the results from different theoretical models about the neutron-proton effective mass splitting, there is almost a consensus reached on the fact that, neutron effective mass in neutron rich matter will go over the proton effective mass. For a given splitting of $\varepsilon_{ex}$ into $\varepsilon_{ex}^l$ and $\varepsilon_{ex}^{ul}$, the splitting of other two strength parameters $\varepsilon_{0}$ and $\varepsilon_{\gamma}$ into like and unlike components requires the knowledge of the zero temperature nuclear symmetry energy  $E_s(\rho_0)$ and its slope $E_s^{\prime}(\rho_0)=\rho\frac{dE_s(\rho)}{d\rho}|_{\rho=\rho_0}=\frac{1}{3}L(\rho_0)$ at saturation density. During the last one or two decades, there have been a lot of experimental and theoretical efforts to constrain these parameters. However, amidst all these efforts, still there exist uncertainties in the values of $E_s(\rho_0)$ and $L(\rho_0)$ \cite{Baran2005, Steiner2005, Li2008, Tsang2012, Lattimer2012, Horowicz2014, Dutra2012, Oertel2017, Wang2015a}. While the $E_s(\rho_0)$ is constrained to a narrow range of  $E_s(\rho_0)=32\pm 4$ MeV, there is a large uncertainty in the values of $L(\rho_0)$. Different theoretical models predict similar $E_s(\rho)$ at saturation density but widely differ in the values of $L(\rho_0)$ and therefore predict quite different density dependence of NSE. The slope parameter has been constrained recently in the range $L(\rho_0)=58.7\pm 28.1$ MeV  from an averaging over a good number of experimental and observational data \cite{Oertel2017}. Constraints on the NSE have been obtained in recent times from the properties of finite nuclei such as the nuclear binding energy \cite{Moller2012, Lattimer2013, Dan2009}, neutron skin thickness \cite{Klos2007, Maza2015, Tamii2011, Rossi2013, Chen2010, Vinas2014, Mondal2016}, electric response ( dipole polarisability)\cite{Zhang2014}. Properties of finite nuclei provide stringent constraints on $E_s(\rho)$ and $L(\rho)$ at subsaturation densities. Using different microscopic and phenomenological models, Fuchs and Wolter  have obtained the NSE at a subsaturation density around $\rho_c\simeq 0.6\rho_0$ to be $E_s(\rho_c) \simeq 24$ MeV \cite{Fuchs2006}. Brown used the properties of doubly magic nuclei to constrain the nuclear equation of state at a density of $\rho_c=0.1$ fm$^{-3}$ \cite{Brown2013}. From an analysis of the binding energy difference of heavy isotope pairs,  Zhang and Chen  have obtained a tighter constraint on the symmetry energy at subsaturation density $\rho_c=0.11$ fm$^{-3}$ i.e. $E_s(\rho_c)=26.65\pm 0.20$ MeV \cite{Zhang2013}. It is worth to mention here that, the average density of heavy nuclei such as $^{208}$Pb is around $0.11$ fm$^{-3}$. On the other hand, density dependence of NSE at high density is quite uncertain. However there have been a significant progress in understanding the high density behaviour of NSE from terrestrial nuclear laboratories and astrophysical observations. The upper limit of tidal deformability for canonical neutron stars $\Lambda_{1,4}=580$ extracted from GW170817 by LIGO and VIRGO collaborators \cite{Abott2018} has allowed to constrain the high density nuclear equation of state to a narrow range. From the analysis of these observations for a constant maximum mass of $M_{max}=2.01 M_{\astrosun}$ and radius $R_{1,4}=12.83$, the high density behaviour of NSE has been constrained at a density $\rho=2\rho_0$  to be $E_s(2\rho_0)= 46.9 \pm 10.1$ MeV \cite{Zhang2019, Li2019}. Using the extracted upper limit on the tidal deformability of the gravitational waves from GW170817, Tong et al. have also obtained a constraint on the NSE at twice saturation density as $E_s(2\rho_0)\leq 60.7 \pm 10.9$ MeV \cite{Tong2020}. Bayesian analysis of the radii of canonical Neutron Stars predicts $E_s(2\rho_0)=39.2^{+12.1}_{-8.2}$ MeV \cite{Xie2019}. Zhang et al. constrained the NSE from an analysis of heavy ion collision data, neutron skin of $^{208}$Pb, tidal deformability and maximum mass of neutron stars and obtained $E_s(2\rho_0)$ in the range $35-55$ MeV \cite{Zhang2020}. More accurate determination of the millisecond pulsars from pulse-profile modelling of accretion hot spots by NASA's \textit{Neutron Star Interior Composition Explorer} (NICER) mission is expected to advance our understanding of the EoS of cold dense matter \cite{nicer1, nicer2, nicer3, nicer4, nicer5}.  It appears that the NICER results and the newly discovered heavy neutron star PSR J0740+6620 \cite{Cromartie2019} support stiffer rather than softer choices of the asymmetric nuclear matter EoS. In order to constrain our interaction parameters so as to provide a good description of the nuclear symmetry density both at a subsaturation cross density as well as at a reasonably high density (as applicable to neutron stars), in the present work, we have varied the slope parameter for a given $E_s(\rho_0)$ within its acceptable range to obtain the $E_s(\rho_c)=26.65 \pm 0.2$ fm$^{-3}$. The slope parameter that fixes the constrained $E_s(\rho_c)$ is then checked so as to predict  the NSE at $2\rho_0$ close to the limit $E_s(2\rho_0)= 46.9 \pm 10.1$ MeV. In this procedure, we obtained the splitting of the strength parameters $\varepsilon_{0}$ and $\varepsilon_{\gamma}$ into like and unlike components. We have considered three different values of $E_s(\rho_0)$ namely $E_s(\rho_0)=33, 34$ and $35$ MeV and obtained the corresponding splitted parameter sets. The parameters of the given finite range SEI as constrained above for the EoS of PNM are given in Table II. The slope parameter for all the sets lies in the acceptable range $44\leq L(\rho_0)\leq 66$ MeV. For all the sets of nuclear equation of state, the values of $K_{sym}(\rho_0)$ lie in the range $-201\leq K_{sym}(\rho_0)\leq -126$ MeV. The NSE at $2\rho_0$ lies in the range $38.65 \leq E_s(2\rho_0)\leq 52.15$ MeV. These values are well within those predicted from terrestrial and astrophysical observations \cite{Zhang2019, Li2019, Zhang2020}. For completeness, we have shown the EoSs for SNM and PNM for the three different parameter sets of the finite range effective interaction in  Figure 1.

\subsection{Density dependence of Nuclear Symmetry Energy}
In Figure 2, we have shown the density dependence of NSE for three different sets of parameters. Also, we have plotted the NSE as obtained from a comparison of ASY-EOS data concerning the elliptic-flow ratio of neutrons with respect to charged particles with ultra relativistic quantum molecular dynamics (UrQMD) transport model \cite{Russotto2016}. In that study,  Russotto et al. have used parametrized form of NSE 
\begin{equation}
E_s(\rho)~\text{[MeV]}=12 \left(\frac{\rho}{\rho_0}\right)^{2/3}+ 22\left(\frac{\rho}{\rho_0}\right)^{\sigma}\label{eq:13},
\end{equation}
and constrained the exponent as $\sigma=0.72\pm 0.19$ \cite{Russotto2016}. This is an improvement over a similar study from FOPI-LAND experiment that suggested a moderately soft to linear NSE characterised by an exponent $\sigma=0.9\pm 0.4$ \cite{Russotto2011}. This study ruled out the supersoft scenarios for NSE. It is worth to mention here that, the supersoft scenarios for NSE were ruled out after the parameter dependence of the ASY-EOS result was investigated by Cozma et al. \cite{Cozma2013}. In Fig. 2,  we have also included the behaviour of NSE at subnormal densities as obtained from different studies \cite{Horowicz2014, Tsang2009}. Our constructed sets of NSE are quite compatible with these results at low density region. However, in the high density region beyond the normal nuclear matter density, our results for the sets with $E_s(\rho_0)= 33$ and $34$ MeV pass below the predicted results from the UrQMD study of ASY-EOS data \cite{Russotto2016}. Only the set with $E_s(\rho_0)=35$ MeV passes through the shaded region (refer to Fig. 2) and therefore may comprise a more suitable equation of state for neutron-rich asymmetric nuclear matter. At normal nuclear matter density, our NSE for set III goes above the ASY-EOS data because of the fact that the parametrized form used in Ref.\cite{Russotto2016} has a sharp value of $E_s(\rho_0)=34$ MeV. Regarding the density dependence behaviour, the NSE for the other two sets can not simply be ruled out on the basis of the ASY-EOS data. Recently, Zhou and Chen \cite{Zhou2019}, on the basis of the discovery of millisecond PSR J0740+6620 with Mass $2.14^{+0.10}_{-0.09}$ M$_{\astrosun}$ \cite{Cromartie2019} together with the data of finite nuclei ruled out super soft NSE that becomes negative at suprasaturation densities in neutron stars. On that basis, the set II with $E_s(\rho_0)=34$ MeV will pass the test. The set I with $E_s(\rho_0)=33$ MeV can be viable according to the constraints of Refs. \cite{Zhang2019, Li2019}. 

\begin{figure}
\includegraphics[width=0.7\textwidth]{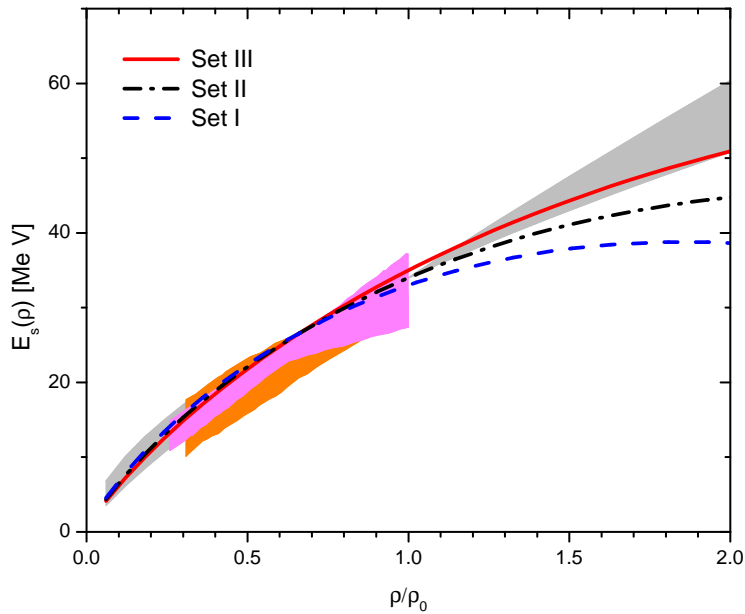}
\caption{Nuclear symmetry energy $E_s(\rho)$ as a function of the reduced density $\frac{\rho}{\rho_0}$. The results at subnormal densities of Ref.\cite{Horowicz2014, Tsang2009} and the UrQMD results (shaded grey region) of ASY-EOS data of Ref. \cite{Russotto2016} are also shown in the figure. The shaded region in orange is for the HIC results of Sn+Sn reaction and the pink region is for the results of IAS. }
\end{figure}

In order to understand the density dependence of NSE as calculated from the finite range SEI, we can split it into its kinetic contribution and the contribution from the interaction part
\begin{equation}
E_s(\rho)=E_s^{kin}(\rho)+E_s^{pot}(\rho)\label{eq:14},
\end{equation}
where
\begin{equation}
E_s^{kin}(\rho)=\frac{3Mc^2}{8}\left[ \frac{2x_nu_n^3-x_nu_f-ln\left(x_n+u_n\right)}{x_n^3}-\frac{2x_fu_f^3-x_fu_f-ln\left(x_f+u_f\right)}{x_f^3}\right]\label{eq:15},
\end{equation}
and 
\begin{eqnarray}
E_s^{pot}(\rho)= \frac{(\varepsilon_0^l-\varepsilon_0)}{2}\frac{\rho}{\rho_0}+\frac{(\varepsilon_{\gamma}^l-\varepsilon_{\gamma})}{2}\frac{\rho}{\rho_0^{\gamma+1}}\left(\frac{\rho}{1+b\rho}\right)^{\gamma}+\frac{[\varepsilon_{ex}^lJ_{n}(\rho)-\varepsilon_{ex}J_{0}(\rho)]}{2}\frac{\rho}{\rho_0}\label{eq:16}.
\end{eqnarray}
In Figure 3, the kinetic and potential contributions to the NSE are shown as function of the reduced nuclear density $\frac{\rho}{\rho_0}$. It is obvious that, the kinetic part has a positive contribution behaving as $\left(\frac{\rho}{\rho_0}\right)^{2/3}$. But the potential contribution depends on the value of $E_s(\rho_0)$ and $L(\rho_0)$. For a given pair of $E_s(\rho_0)$ and $L(\rho_0)$, the potential part $E_s^{pot}(\rho)$ increases to attain a peak and then decreases with an increase in the reduced nuclear density. The peak value and the corresponding reduced density $\left(\frac{\rho}{\rho_0}\right)_{max}$ increase with the increase in the values of $E_s(\rho_0)$ and $L(\rho_0)$. Consequently, the $E_s^{pot}(\rho)$ with low values of $E_s(\rho_0)$ and $L(\rho_0)$ quickly becomes negative and thereby makes the NSE soft or supersoft. Since the exchange strength parameters $\varepsilon_{ex}$ and $\varepsilon_{ex}^{l}$ are fixed for all the sets, the contribution coming from the exchange term remains the same. In fact, this exchange contribution to NSE increases with an increase in the nuclear matter density. One can note that, the contribution from the term involving the parameters $\varepsilon_0$ and $\varepsilon_0^l$ is positive and the contribution from the term involving the parameters $\varepsilon_{\gamma}$ and $\varepsilon_{\gamma}^l$ is negative. This negative contribution decreases as we increase the $E_s(\rho_0)$ value. Consequently, the potential contribution to NSE becomes more and more negative after certain nuclear matter density leading to a softer NSE.

\begin{figure}
\includegraphics[width=0.7\textwidth]{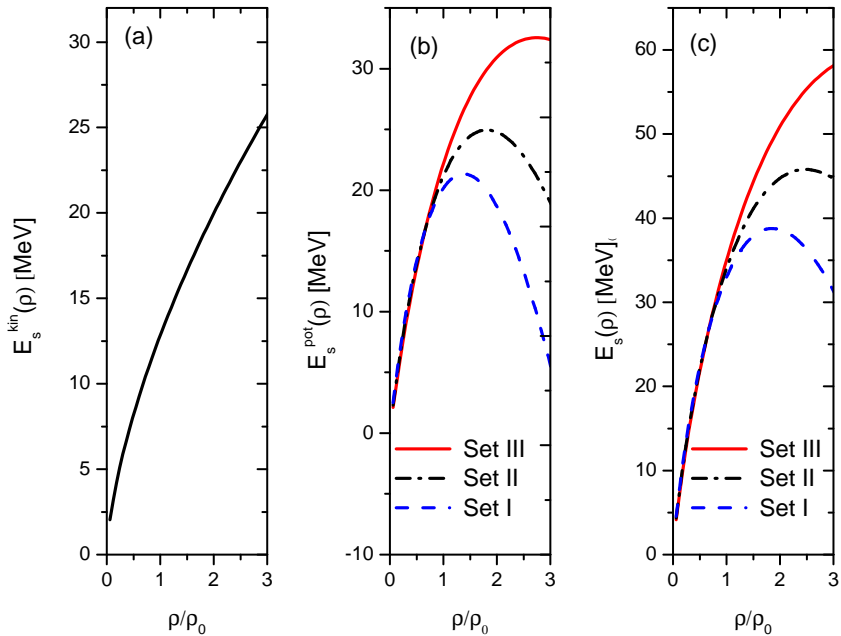}
\caption{Components of Nuclear symmetry energy $E_s(\rho)$ as a function of the reduced density $\frac{\rho}{\rho_0}$. In the extreme left panel we have shown the kinetic energy contribution to the NSE. In the middle panel, the contribution coming from the interaction part is shown and in the right panel, the total NSE is plotted.}
\end{figure}

The slope parameter $L(\rho_0)=3E_s^{\prime}(\rho_0)$  as obtained from the fitting procedure can be correlated with the 
$E_s(\rho_0)$. In fact these two parameters decide the density dependence of NSE and are highly correlated. From different mean field models, it is found that, these two parameters are linearly correlated. Roca-Maza et al. have obtained a correlation between $E_s(\rho_0)$ and $L(\rho_0)$ from the experimentally measured values $19.6\pm 0.6$ fm$^{3}$ of the dipole polarisability of $^{208}$Pb as $E_s(\rho_0)=(24.5\pm 0.8)+(0.168\pm 0.007)L(\rho_0)$ \cite{Maza2015}. Lattimer and Steiner \cite{Lattimer2014} have obtained a similar but slightly different correlation relation from the experimental value of the dipole polarisability of $^{208}$Pb as reported by Tamii et al. \cite{Tamii2011}. In Figure 4, we have shown these experimentally extracted correlation of $E_s(\rho_0)$ and $L(\rho_0)$. In the figure, we have also shown the correlation from IAS results as extracted from Ref. \cite{Lattimer2014}.  The red dots are the results of the present work as constrained from the fitting procedure of the interaction parameters of the finite range effective interaction. At a given $E_s(\rho_0)$, we have obtained three different values of $L(\rho_0)$ corresponding to the fitting of the NSE to the value of $E_s(\rho_c)=26.65\pm 0.2$ MeV. The correlation between $E_s(\rho_0)$ and $L(\rho_0)$ from our calculations are more or less similar and compatible to that of Roca-Maza et al. It may be noted from the figure that all the points of our calculation lie in the black hatched region even though the results corresponding to $E_s(\rho_0)=35$ MeV overlap with the region extracted by Lattimer and Steiner.

\begin{figure}[t]
\includegraphics[width=0.7\textwidth]{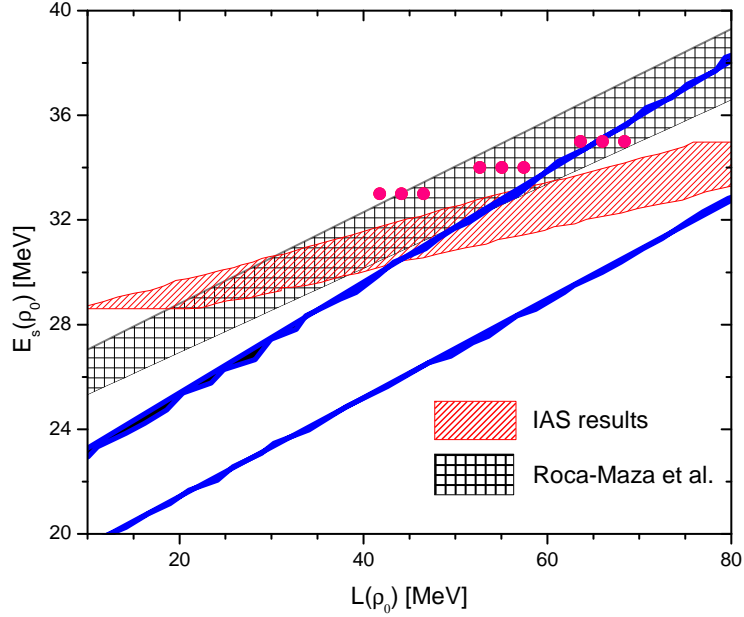}
\caption{Nuclear symmetry energy $E_s(\rho)$ as a function of  $L(\rho_0)$ as extracted from different works. The black hatched region represents the results of Roca-Maza et al. \cite{Maza2015}. The region between the two thick blue lines represent the results of Lattimer and Steiner \cite{Lattimer2014} . The red ruled area is the correlation from IAS results as extracted from Ref.\cite{Lattimer2014}. The red dots are the calculated values from the present work.}
\end{figure}

In order to have an idea of the behaviour of the symmetry energy at a reference density $\rho_c < \rho_0$, we may expand the NSE around $E_s(\rho_c)$ as 
\begin{equation}
E_s(\rho)=E_s(\rho_c)+L(\rho_c)\varepsilon+\frac{K_{sym}(\rho_c)}{2!}\varepsilon^2+\mathcal{O}(\varepsilon^3)\label{eq:17},
\end{equation}
where $\varepsilon=\frac{\rho-\rho_c}{3\rho_c}$. $L(\rho_c)=3\rho_c\frac{dE_s(\rho)}{d\rho}|_{\rho=\rho_c}$ is the density slope parameter and $K_{sym}(\rho_c)=9\rho_c^2\frac{d^2E_s(\rho)}{d\rho^2}|_{\rho=\rho_c}$ is the curvature parameter at the reference density $\rho_c$. Knowledge of $L(\rho_c)$ besides being an important quantity in determining the density dependence of the NSE at low density region, plays an important role in determining the neutron skin thickness of heavy nuclei. In view of this, more precise constraints on $L(\rho_c)$ will help a lot in the understanding of the nuclear equation of state. We have plotted the NSE as a function of the parameter $\varepsilon$ in Figure 5 for the three sets assuming the reference density as the central density for $^{208}$Pb. In the figure, the NSE of Russotto et al. \cite{Russotto2016} with a parametrized form as in Eq.\eqref{eq:13} and the extracted value of the exponent as $\sigma=0.72$ is also shown for comparison. The values of the parameters $L(\rho_c)$ and $K_{sym}(\rho_c)$ have also been calculated from the expression of the NSE. The calculated values of $L(\rho_c)$ and $K_{sym}(\rho_c)$ are given in the Table III. Recently Zhang and Chen \cite{Zhang2014} have extracted the value of $L(\rho_c)$ from the analysis of the electric dipole polarisability in $^{208}$Pb as $L(\rho_c)=47.3\pm 7.8 $ MeV. From the present calculations, the $L(\rho_c)$ are predicted to be $46.69\leq L(\rho_c) \leq 55.63$ MeV which are well within the extracted values  of Zhang and Chen \cite{Zhang2014}. Using the finite range effective interaction, we have obtained $K_{sym}(\rho_c)$ in the range $-120.69 \leq K_{sym}(\rho_c) \leq -82.8$ MeV. It is worth to mention here that, while empirical constraints on $L(\rho_c)$ are available in literature there are no such constraints on the parameter $K_{sym}(\rho_c)$.

\begin{figure}
\includegraphics[width=0.7\textwidth]{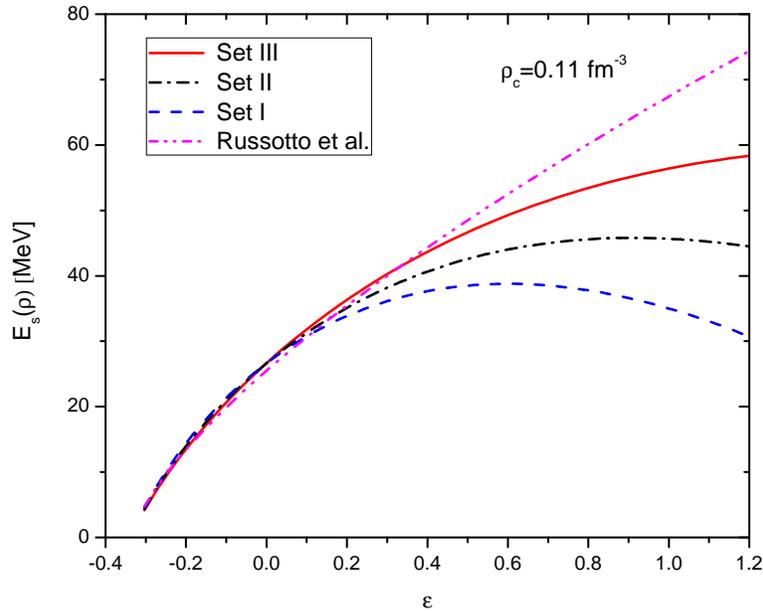}
\caption{Nuclear symmetry energy $E_s(\rho)$ as a function of  $\varepsilon=\frac{\rho-\rho_c}{3\rho_c}$. The results of Ref. \cite{Russotto2016} are also shown in the figure for comparison. }
\end{figure}

\begin{table}
\caption{Prediction of different isovector indicators from the finite range effective interaction.}
\centering
\begin{tabular}{c|c|c|c|c|c|c|c}
\hline
\hline
Parameter sets & $L(\rho_0)$ & $L(\rho_c)$ & $K_{sym}(\rho_0)$&$K_{sym}(\rho_c)$ & $a_{ss}$ & $a_{cs}$& $Q$\\
	   & (MeV)	      & (MeV)       & (MeV)& (MeV)           &(MeV)	               & (MeV)    &(MeV)\\
\hline
Set I      & 44.16         &   46.69    & -201.32 &-120.69   & 43.88 & 36.46 & 52.59 \\

Set II	   & 55.05         &   51.15    & -168.86 &-101.82   & 52.44 & 51.17 & 46.81 \\

Set III    & 66.0          &   55.63    & -136.16 &-82.80    & 61.05 & 65.97 & 42.42\\

\hline
\end{tabular}
\end{table}

\section{Symmetry energy coefficient}
To understand the behaviour of nuclear symmetry energy in nuclear masses, it is essential to have a correlation of the symmetry energy coefficient in finite nuclei $a_{sym}(A)$ and the NSE $E_s{(\rho)}$. Within the droplet model (DM), Centelles et al. \cite{Centelles2009, Warda2009} have found that the symmetry energy coefficient $a_{sym}(A)$ of finite nuclei with mass number $A$ is equal to the nuclear symmetry energy, $E_s(\rho_A)$ calculated at the central density $\rho_A$ of the nucleus i.e $a_{sym}(A)\simeq E_s(\rho_A)$. They have shown that, this relation not only works for heavy nuclei such as $^{208}$Pb but also holds for medium mass nuclei  \cite{Centelles2009}. Such a relation is helpful in providing a direct correlation of the isospin observables of finite nuclei with $E_s{(\rho)}$ at subnormal densities.

The symmetry energy coefficient $a_{sym}(A)$ in finite nuclei can also be expressed as \cite{Centelles2009}
\begin{equation}
a_{sym}(A)=\frac{E_s(\rho_0)}{1+x_A}\label{eq:18},
\end{equation}
where 
\begin{equation}
x_A=\frac{9}{4}\frac{E_s(\rho_0)}{Q}A^{-1/3}\label{eq:19}.
\end{equation}
$Q$ is the surface stiffness parameter that measures the resistance of the nucleus against separation of neutrons from protons to form a skin. This quantity is directly related to the nuclear surface symmetry energy \cite{Myers1980, Dan2009}. Usually $Q$ is obtained from semi-infinite nuclear matter calculations \cite{Brack1985, Centelles1998}. The surface stiffness parameter can be calculated from the expression \cite{Treiner1986, Chen2011}
\begin{equation}
Q=\frac{9}{4}\left(L(\rho_0)-\frac{K_{sym}(\rho_0)}{12}\right)^{-1}E_s^2(\rho_0)\label{eq:20}.
\end{equation}
However, in the present work, we have used the expression
\begin{equation}
Q=\frac{9}{4}\left(\frac{E_s(\rho_0)}{a_{sym}(A)}-1\right)^{-1}E_s(\rho_0)A^{-1/3}\label{eq:21}.
\end{equation} 
The calculated values of the surface stiffness coefficient are given in Table III. One can note that, with an increase in the values of $L(\rho_0)$ and $E_s(\rho_0)$, the value of $Q$ decreases.

Assuming the relation $a_{sym}(A)\simeq E_s{(\rho_A)}$ to be valid in a wide range of nuclear mass, we have calculated the symmetry energy coefficient for the three sets of nuclear EoSs. For this purpose we have used the density parametrization
\begin{equation}
\rho_A=\frac{\rho_0}{1+\kappa A^{-1/3}}\label{eq:22},
\end{equation}
where the constant $\kappa$ is fixed to reproduce the density of $^{208}$Pb. In Figure 6, the reference density $\rho_A$ as calculated from Eq. \eqref{eq:22} is shown as a function of mass number $A$. The reference density of heavy nuclei are usually considered in literature as close to $0.1$ fm$^{-3}$. The same has been reflected in the figure. The reference density increases with $A$ and almost saturates for large nuclear masses. In Figure 7, the symmetry energy coefficient of finite nuclei (as obtained in an approximate manner) is plotted as a function of the mass number for the three different sets of interaction  parameters. In the plot, we have extrapolated the results to large values of nuclear mass $A=1000$ to get an idea of the asymptotic value of $a_{sym}(A)$. As the mass number increases, $a_{sym}(A)$ increases slowly to reach its asymptotic value. In the figure, we have also shown the $a_{sym}(A)$ as calculated by using SLy4 interaction \cite{Chabanat1998, Wang2015}. Even though the behaviour of the curve for SLy4 is the same as that of the curves for finite range effective interaction SEI, it lies below the SEI curves. The reason may lie in the fact that, the finite range interaction parameters are fitted in such a manner to reproduce the empirically constrained value of $E_s(\rho_c)=26.65\pm 0.2$ MeV for $^{208}$Pb \cite{Zhang2013} whereas for the SLy4 interaction $a_{sym}(208)$ is around $22.90~ MeV$. It can be observed that, the three sets namely Set I, Set II and Set III predict almost similar values for $a_{sym}(A)$ with a little spread both at the high mass and low mass region. 

\begin{figure}[t]
\minipage{0.49\textwidth}
\centering
\includegraphics[width=85mm]{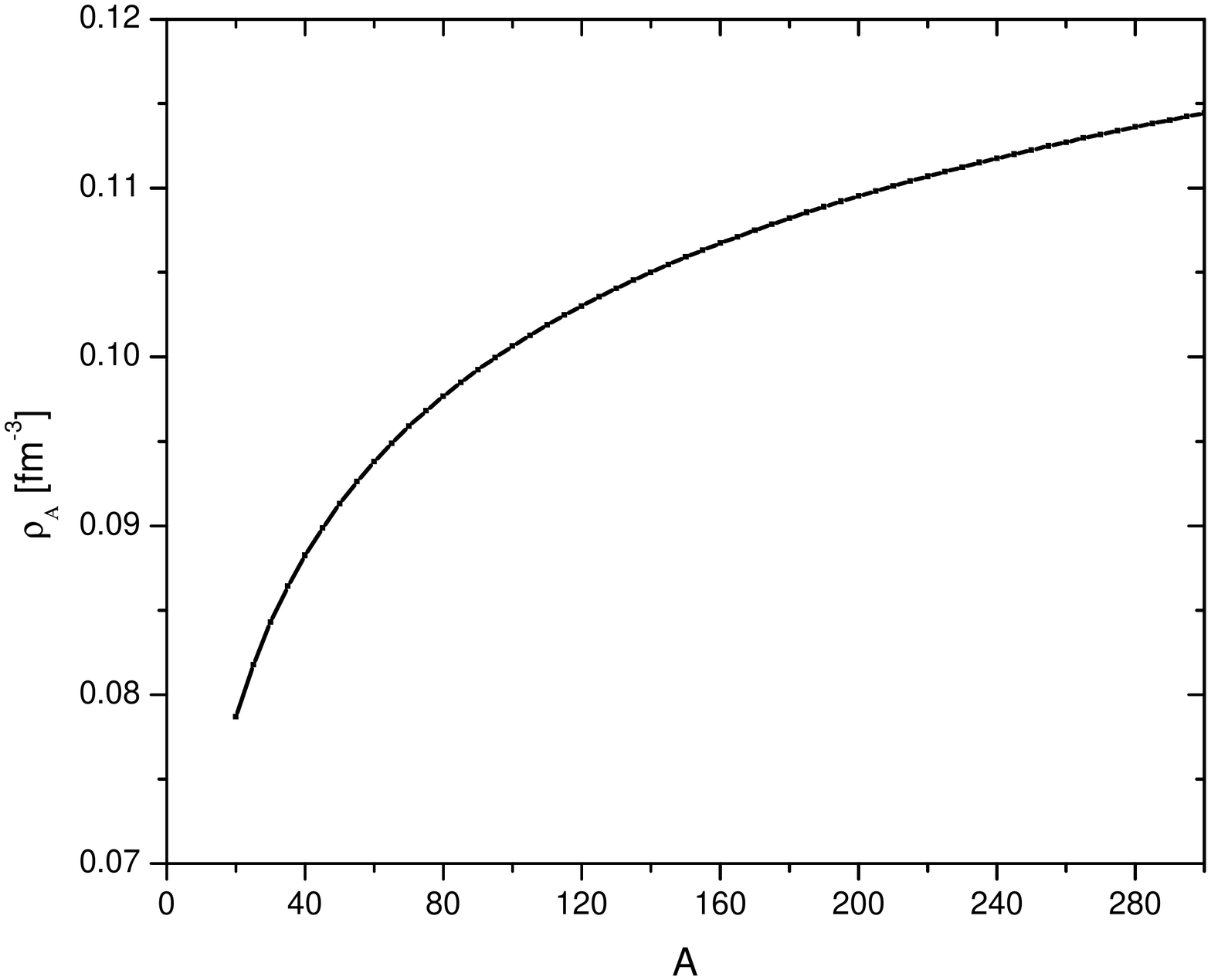}
\caption{Reference density $\rho_A$ as a function of mass number $A$. The reference density for $^{208}$Pb is taken as $0.11$ fm$^{-3}$. } 
\endminipage
\hfill
\minipage{0.49\textwidth}
\includegraphics[width=85mm]{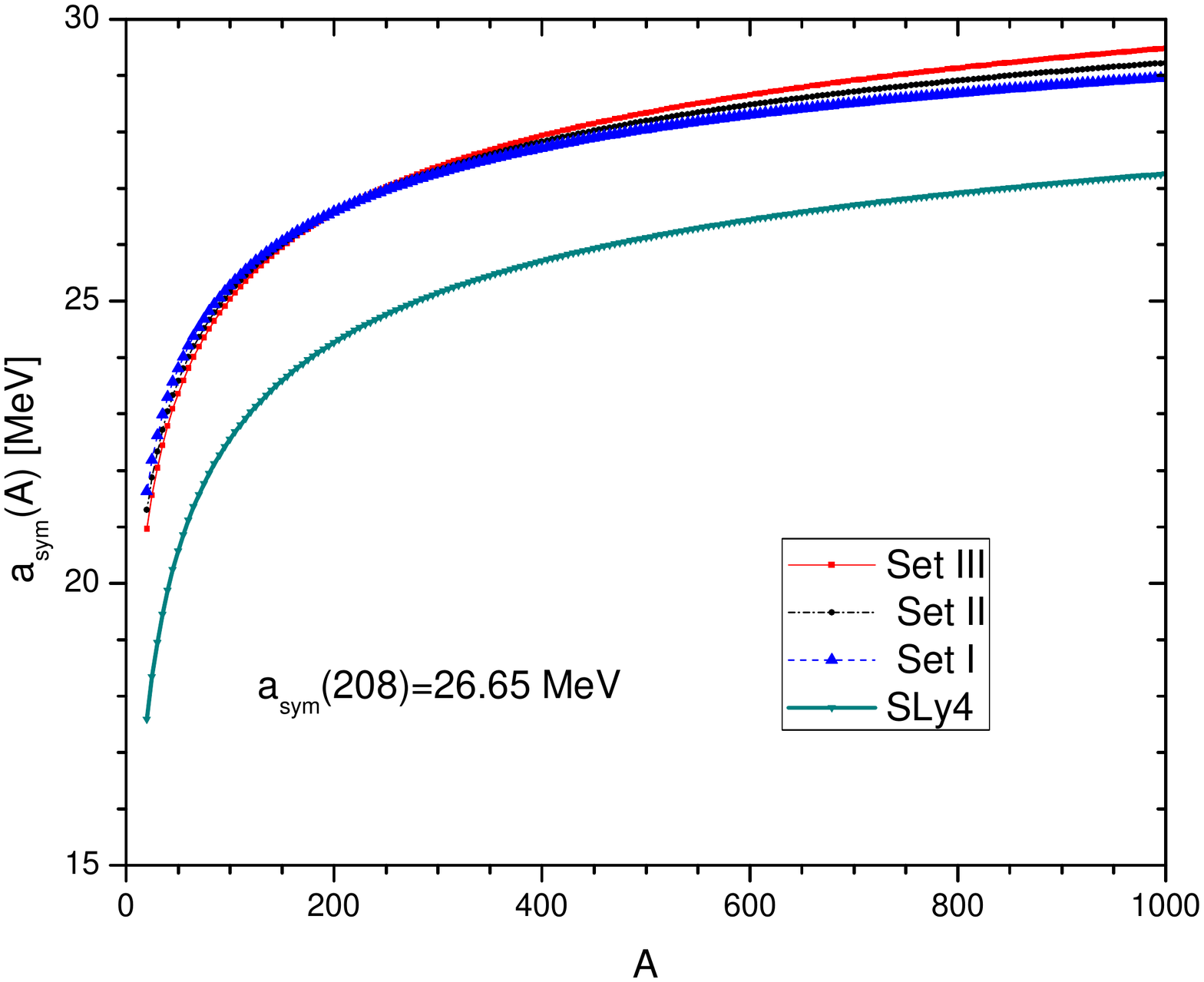}
\caption{Symmetry energy coefficient $a_{sym}(A)$ as a function of mass number $A$ for the three EoSs. The results of SLy4 interaction \cite{Chabanat1998} are also shown in the graph for comparison. }
\endminipage
\end{figure}

In a leptodermous expansion within liquid drop model, the symmetry energy coefficient can be expanded in powers of $A^{-1/3}$ as
\begin{equation}
a_{sym}(A)=E_s(\rho_0)-a_{ss}A^{-1/3}+a_{cs}A^{-2/3}\label{eq:23},
\end{equation}
where $a_{ss}$ is the coefficient of the surface symmetry term and $a_{cs}$ is the coefficient of the curvature symmetry term. We have fitted the $a_{sym}(A)$ as calculated from Set I, Set II and Set III to Eq.\eqref{eq:23} and extracted the coefficients $a_{ss}$ and $a_{cs}$. The values of these coefficients are given in Table III. One can note that, the values of $a_{ss}$ and $a_{cs}$ increase with an increase in the value of $L(\rho_c)$.

\begin{figure}
\includegraphics[width=0.7\textwidth]{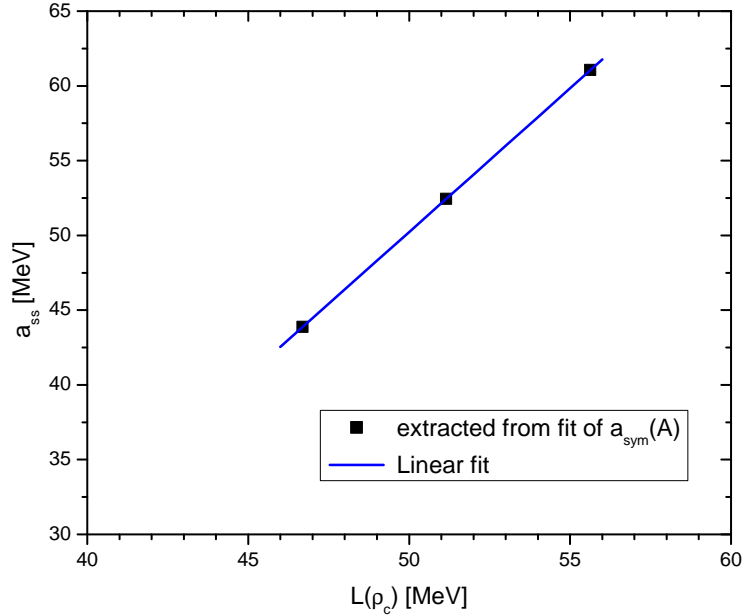}
\caption{Surface symmetry coefficient $a_{ss}$ as extracted from the leptodermous expansion of $a_{sym}(A)$ as a function of the density slope parameter $L(\rho_c)$ at reference density. Here the $L(\rho_c)$ values are considered for $^{208}$Pb as given in Table III. The plot also shows a linear fit between $a_{ss}$ and $L(\rho_c)$. }
\end{figure}

Keeping up to first order in $A^{-1/3}$ in Eq. \eqref{eq:23}, we can express the surface symmetry term as
\begin{equation}
a_{ss}=\left[E_s(\rho_0)-a_{sym}(A)\right]A^{1/3}\label{eq:24}.
\end{equation}
Substituting for $a_{sym}(A)$ from Eq.\eqref{eq:17} in Eq. \eqref{eq:23}, we obtain
\begin{equation}
a_{ss}=\left[E_s(\rho_0)-E_s(\rho_c)-L(\rho_c)\varepsilon\right]A^{1/3}\label{eq:25}.
\end{equation}
Here the relation $a_{sym}(A)\simeq E_s{(\rho_A)}$ is assumed and the expansion in Eq.\eqref{eq:17} is truncated after  first order term in $\varepsilon$. This equation \eqref{eq:25} clearly indicates a linear relation between the surface symmetry term and the density slope parameter at a reference density. In Figure 8, we have shown the linear correlation between the extracted $a_{ss}$ and $L(\rho_c)$. A linear fit of the data points provides the relationship
\begin{equation}
a_{ss}=-45.872+ 1.922L(\rho_c)\label{eq:26}.
\end{equation}

\section{Neutron skin thickness of $^{208}Pb$}

The neutron skin thickness (NST) is defined as the difference between the rms radii for the density distributions of the neutrons and protons in the nucleus,
\begin{equation}
\varDelta r_{np}=\left\langle r^2\right\rangle_n^{1/2}-\left\langle r^2\right\rangle_p^{1/2}\label{eq:27},
\end{equation}
which can be expressed as \cite{Mondal2016}
\begin{equation}
\varDelta r_{np}= \varDelta r_{np}^{bulk}+\varDelta r_{np}^{Coul}+\triangle r_{np}^{surf}\label{eq:28}.
\end{equation}
The bulk part $\varDelta r_{np}^{bulk}=\sqrt{\frac{3}{5}}\left(R_n-R_p\right)$ is proportional to the distance between the neutron and proton radii of uniform sharp distributions. $\varDelta r_{np}^{Coul}=-\sqrt{\frac{3}{5}}\frac{e^2Z}{70E_s(\rho_0)}$ is a correction due to the Coulomb repulsion and $\varDelta r_{np}^{surf}=2.5\sqrt{\frac{3}{5}}\left(\frac{b_n^2}{R_n}-\frac{b_p^2}{R_p}\right)\simeq \sqrt{\frac{3}{5}}\frac{5}{2R}\left(b_n^2-b_p^2\right)$ is a correction due to the difference in the surface widths $b_n$ and $b_p$ of the neutron and proton profiles.

It is well known that, the liquid drop model provides a useful tool to correlate the NST with different NSE parameters. In the present section, we wish to find out the correlations among the neutron skin thickness of $^{208}$Pb and  different isovector parameters of the nuclear symmetry energy. Within the purview of DM \cite{Myers1980} and neglecting the shell correction, the bulk part of NST can be expressed as
\begin{equation}
\varDelta r_{np}^{bulk}=\sqrt{\frac{3}{5}}t \label{eq:29},
\end{equation}
where 
\begin{equation}
t=\frac{3}{2}r_0\frac{E_s(\rho_0)}{Q}\left(\frac{I-I_c}{1+x_A}\right)\label{eq:30}.
\end{equation}
Here $R=r_0A^{1/3}$, $r_0=\left(\frac{4}{3}\pi\rho_0\right)^{-1/3}$ and $I_c=\frac{e^2Z}{20E_s(\rho_0)R}$ is the Coulomb correction to the symmetry energy coefficient.

Substituting the expressions of $x_A$ from Eq. \eqref{eq:18} and $Q$ from Eq. \eqref{eq:21} into the above relation \eqref{eq:30}, we obtain,
\begin{equation}
t=\frac{2}{3}r_0\left[1-\frac{a_{sym}(A)}{E_s(\rho_0)} \right]A^{1/3}(I-I_c)\label{eq:31}.
\end{equation}
Using Eq.\eqref{eq:6} and $a_{sym}(A)=E_s(\rho_A)$, Eq.\eqref{eq:31} can be reduced to
\begin{equation}
t\simeq -2r_0\epsilon_A\beta\left(1+\frac{K_{sym}(\rho_0)}{2L(\rho_0)}\epsilon_A\right)A^{1/3}(I-I_c)\label{eq:32},
\end{equation}
where $\beta=\frac{L(\rho_0)/3}{E_s(\rho_0)}=\frac{E_s^{\prime}(\rho_0)}{E_s(\rho_0)}$ and $\epsilon_A=\frac{\rho_c-\rho_0}{\rho_0}$. Obviously, Eqs.\eqref{eq:31} and \eqref{eq:32} suggest a correlation between the bulk part of the NST in finite nuclei and some isovector indicators such as $1-\frac{a_{sym}(A)}{E_s(\rho_0)}, \beta $ and $\frac{K_{sym}(\rho_0)}{E_s(\rho_0)}$. The close correlations among different isovector observables in finite nuclei with symmetry energy parameters have been studied and are reported in literature \cite{Warda2009, Centelles2009, Centelles2010, Vinas2014, Mondal2016, Raduta2018, Maza2015, Zhang2013, Zhang2018, Maza2011, Maza2013}. In most of the studies, the correlation between $\varDelta r_{np}$ or its bulk part of a given nucleus with $E_s(\rho_0)/Q$ or $L(\rho_0)$ have been established. In some cases, $\varDelta r_{np}$ has been correlated with $E_s(\rho_0)-a_{sym}(A)$. The bulk part of NST, or more specifically the quantity $t$, has a dominant contribution to the neutron skin thickness and depends on the nuclear symmetry energy parameters. Basing upon the expressions of $t$ in Eqs. \eqref{eq:31} and \eqref{eq:32}, it is more reasonable to correlate $t$ with the isovector indicators such as $1-\frac{a_{sym}(A)}{E_s(\rho_0)}, \beta $ and $\frac{K_{sym}(\rho_0)}{E_s(\rho_0)}$. In the present work, we have considered three different values of $E_s(\rho_0)$ namely $33, 34$ and $35$ MeV and constrained the EoSs so as to reproduce the symmetry energy for $^{208}$Pb in the range $26.65\pm 0.2$ MeV. In view of this, in the correlation plots, we have considered the combined parameter $1-\frac{a_{sym}(A)}{E_s(\rho_0)}$ taking into account the role of $E_s(\rho_0)$. In Figure 9, we plot $\varDelta r_{np}$ for $^{208}$Pb as a function of $1-\frac{a_{sym}(A)}{E_s(\rho_0)}$. In Figure 10, we show the correlation of the quantity $t$ with $1-\frac{a_{sym}(A)}{E_s(\rho_0)}$. In these figures, the results of all the nine EoSs are shown. The shaded bands in the figures depict the predictions from the regression procedure. The respective Pearson correlation coefficients $C_P$ are mentioned in the figures. One may note that, the Pearson correlation coefficient for Fig. 9 is 0.9995 and that for Fig. 10 is 0.99996. In other words, the correlation for the bulk part of the NST is much better as compared to that of the whole of the neutron skin thickness.

\begin{figure}
\minipage{0.49\textwidth}
\centering
\includegraphics[width=85mm]{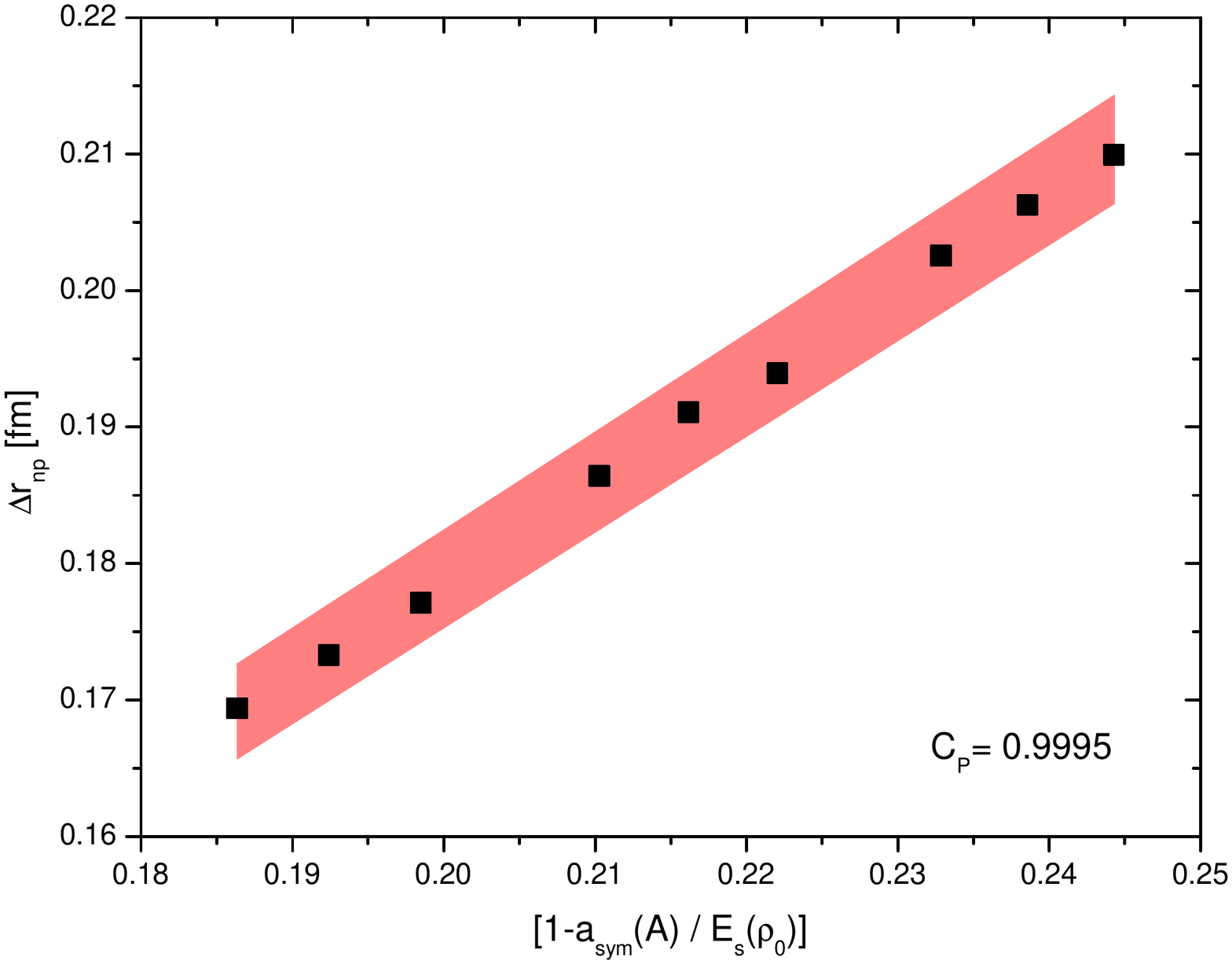}
\caption{The neutron skin thickness in $^{208}$Pb is plotted as a function of the quantity $1-\frac{a_{sym}(A)}{E_s(\rho_0)}$. The shaded region shows the correlation band with a Pearson correlation coefficient $C_P=0.9995$.} 
\endminipage
\hfill
\minipage{0.49\textwidth}
\includegraphics[width=85mm]{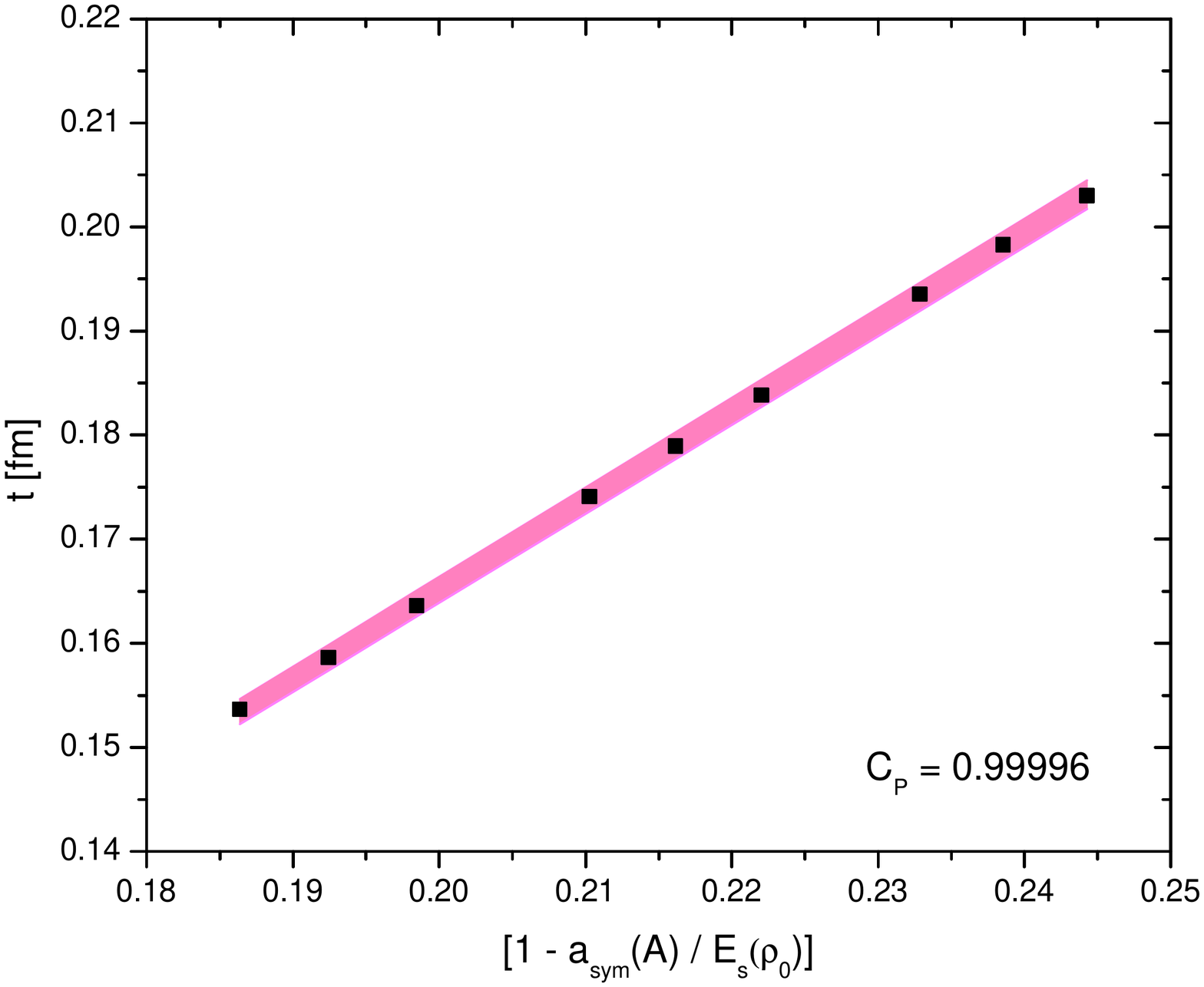}
\caption{ The quantity $t$ for $^{208}$Pb is plotted as a function of the quantity $1-\frac{a_{sym}(A)}{E_s(\rho_0)}$. The shaded region shows the correlation band with a Pearson correlation coefficient $C_P=0.99996$. }
\endminipage\\
\minipage{0.49\textwidth}
\centering
\includegraphics[width=85mm]{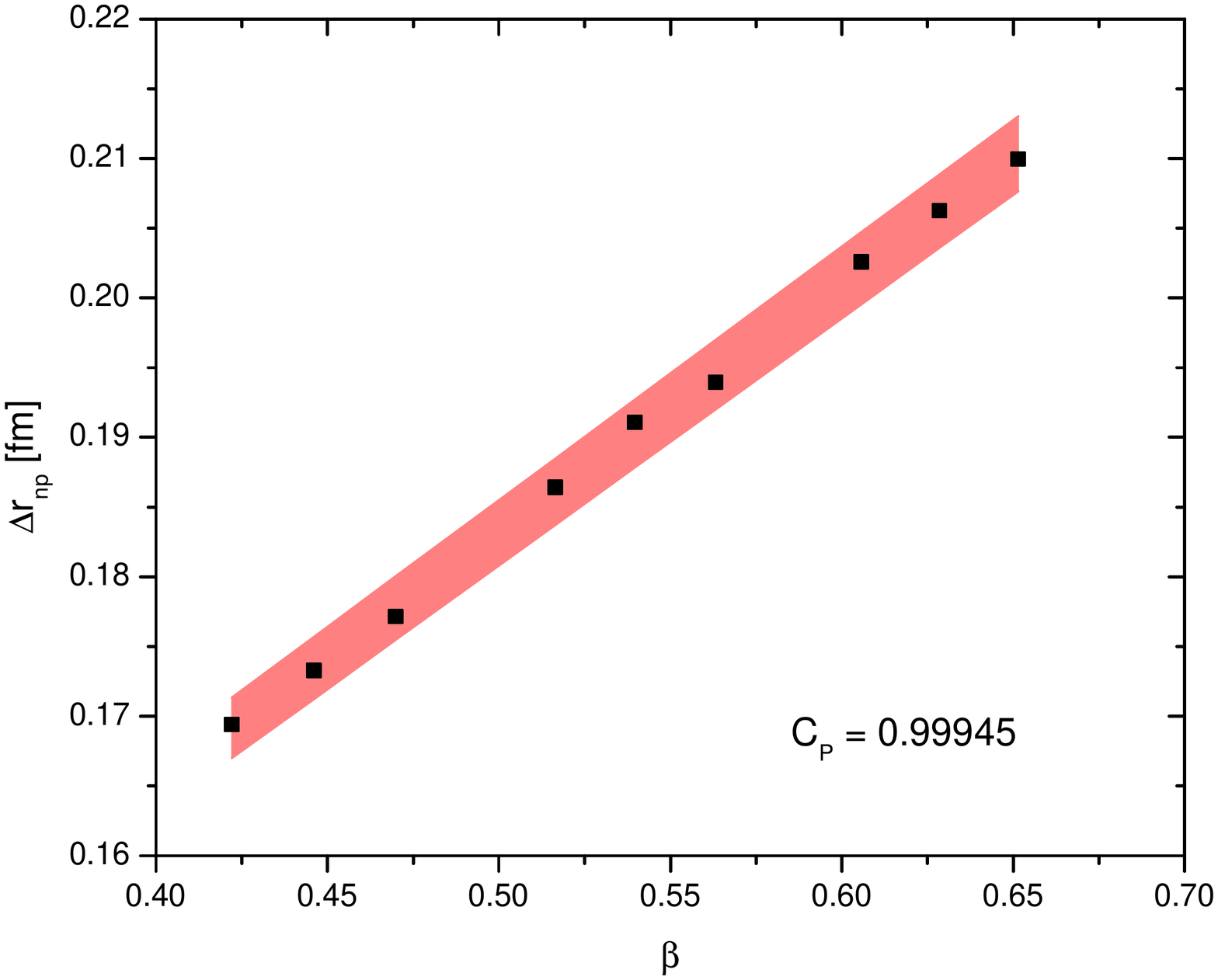}
\caption{The neutron skin thickness in $^{208}$Pb is plotted as a function of the quantity $\beta$. The shaded region shows the correlation band with a Pearson correlation coefficient $C_P=0.99945$. } 
\endminipage
\hfill
\minipage{0.49\textwidth}
\includegraphics[width=85mm]{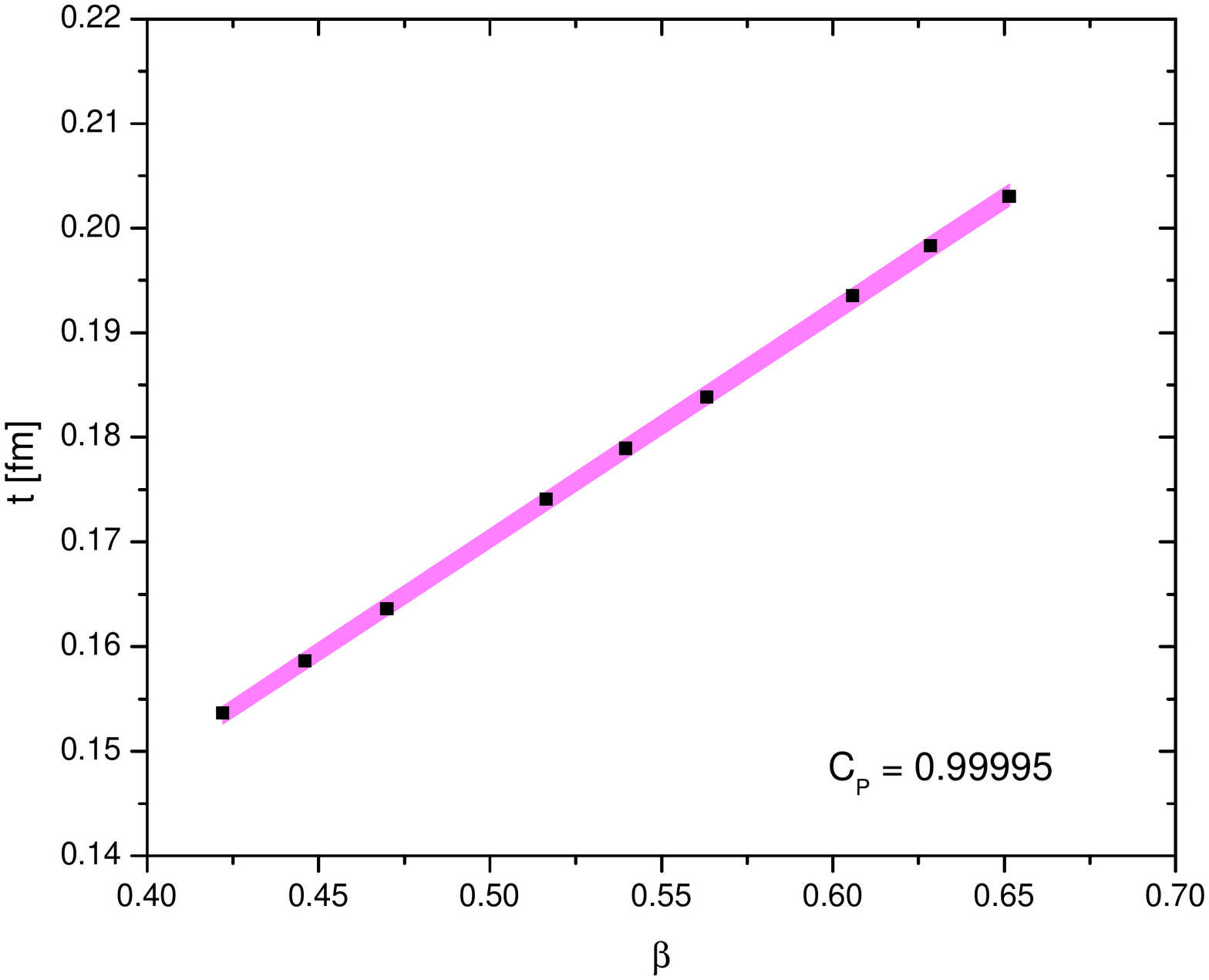}
\caption{The quantity $t$ for $^{208}$Pb is plotted as a function of the quantity $\beta$. The shaded region shows the correlation band with a Pearson correlation coefficient $C_P=0.99995$. }
\endminipage
\end{figure}

One should note that the quantity $t$ is directly proportional to the parameter $\beta=\frac{E_s^{\prime}(\rho_0)}{E_s(\rho_0)}$. Another fact is that,  EoSs with same value of NSE at saturation density may have different slopes. In view of this, we emphasize the correlation between $\varDelta r_{np}$  and $\beta$ instead of finding a compact relation between $\varDelta r_{np}$ and $L(\rho_0)$. For the EoSs employed in this work, $\varDelta r_{np}$ is calculated to be in the range $0.17 - 0.21$ fm for the range of $0.425\leq \beta\leq 0.675$ with a spreading of around $0.04$ fm. In Figures 11 and 12, we have shown respectively the correlation of $\varDelta r_{np}$ and $t$ with the parameter $\beta$. It is interesting to note that, these  quantities are highly correlated in a linear relationship. From a linear fit to the plots, we obtain the relationships
\begin{eqnarray}
\varDelta r_{np}  &=& (0.0934\pm 0.0012)+(0.1796\pm 0.0023)\beta~\text{[fm]},\label{eq:33}\\
t  &=& (0.6196\pm 0.0005)+(0.2167\pm 0.0008)\beta~\text{[fm]}.\label{eq:34}
\end{eqnarray}
The values of $\varDelta r_{np}$ as estimated from the above relationship are in conformity with the predictions from a large sets of relativistic and non relativistic nuclear mean field models \cite{Maza2013}. Usually $L(\rho_0)$ shows a linear relationship with the NST of finite nuclei. Roca-Maza et al. have extracted a linear relationship among $\varDelta r_{np}$ and $L(\rho_0)$ for $^{208}$Pb as $\varDelta r_{np}=0.101+0.00147 L(\rho_0)$ \cite{Maza2011}. Substituting the values of $L(\rho_0)$ in this extracted relation from our work we obtain the NST  in the range $0.16\leq \varDelta r_{np} \leq 0.2$ fm. For completeness, we have plotted the $\varDelta r_{np}$ as a function of $L(\rho_0)$ in Figure 13 and compared our results with that of Ref. \cite{Maza2011}. Results of some mean field calculations are also shown in the figure for comparison. The correlation coefficient in this case is $C_P=0.99943$ a bit less than that of the $\varDelta r_{np}\sim \beta$ plot. A linear fit of the results returns us the relation $\varDelta r_{np}=(0.1066\pm 0.00107)+(0.0015\pm 0.000019)L(\rho_0)$. This expression predicts a bit higher value of the NST as compared to the prediction of the linear fit of  Roca-Maza et al.\cite{Maza2011}. However, our results are well within the predicted regions of Ref.\cite{Maza2011}.

\begin{figure}
\includegraphics[width=0.7\textwidth]{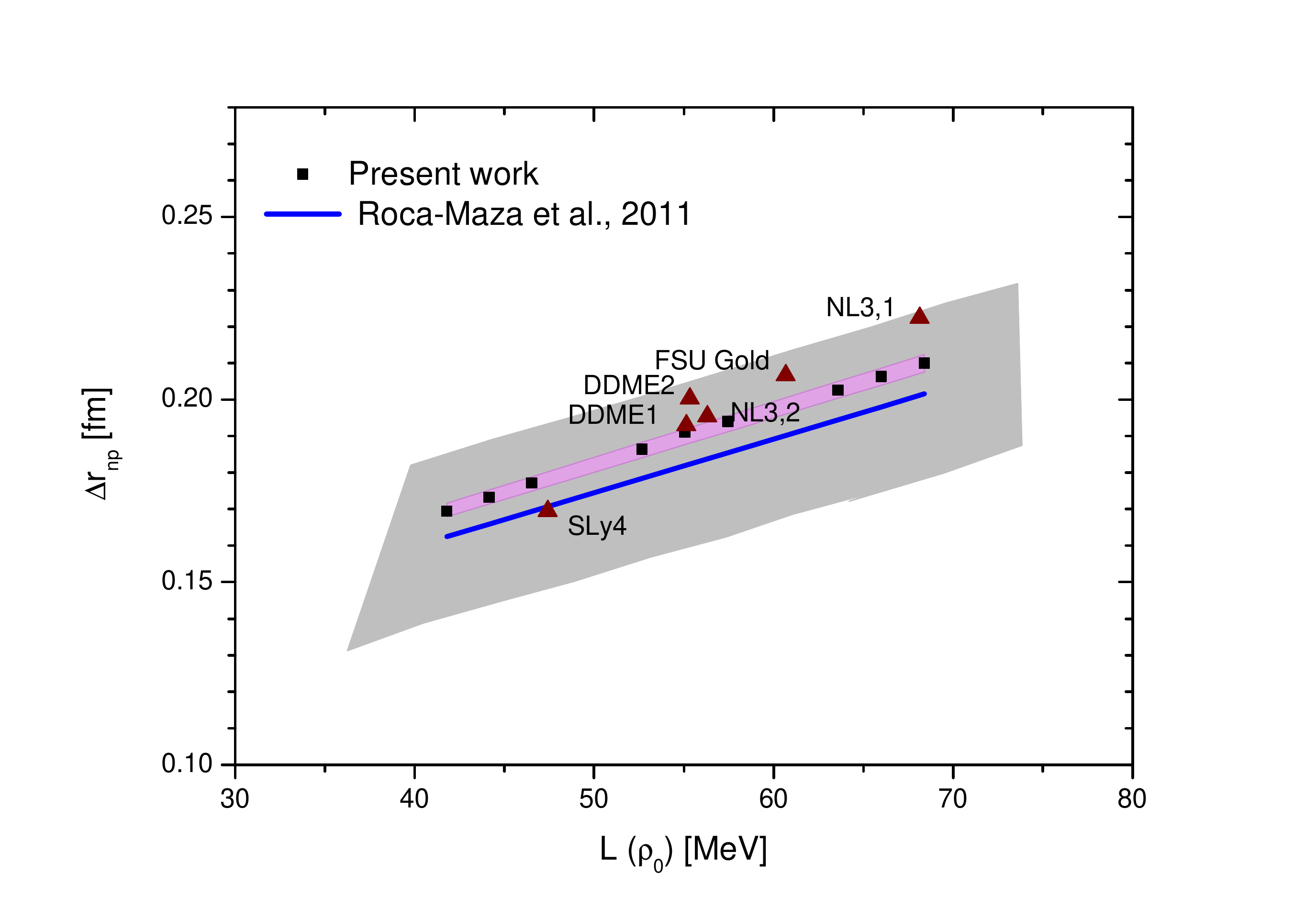}
\caption{The neutron skin thickness in $^{208}$Pb is plotted as a function of the density slope parameter at saturation density. The solid triangles are the extracted results of some of the mean field calculations of Ref.\cite{Maza2011}. The results are compared with the linear fit of the neutron skin thickness in $^{208}$Pb with the density slope parameter as obtained in Ref. \cite{Maza2011}. The grey shaded region corresponds to the predicted region from the regression procedure of Ref.\cite{Maza2011}.}
\end{figure}

We next explore the correlation between the NST and the density slope parameter at a reference density $\rho_c<\rho_0$. Replacing $\rho$ by $\rho_0$ in Eq. \eqref{eq:17} and keeping up to second order in $\varepsilon$, we get
\begin{equation}
\frac{E_s(\rho_0)-a_{sym}(A)}{3E_s(\rho_0)} = \beta^{\prime}\varepsilon\left[1+\frac{1}{2}\frac{K_{sym}(\rho_c)}{L(\rho_c)}\varepsilon\right]\label{eq:35},
\end{equation}
so that the distance between the neutron and proton mean surface locations becomes
\begin{equation}
t=2r_0\varepsilon \beta^{\prime}\left[1+\frac{1}{2}\frac{K_{sym}(\rho_c)}{L(\rho_c)}\varepsilon\right]A^{1/3}(I-I_c)\label{eq:36},
\end{equation}
where $\beta^{\prime}=\frac{L(\rho_c)}{3E_s(\rho_0)}$  and the present value of $\varepsilon$ is $\frac{\rho_0-\rho_c}{3\rho_c}$. Here we have used the fact that $\rho_A\simeq \rho_c<\rho_0$ and $A$ corresponds to the mass number of the nucleus with central density $\rho_A$. Eq. \eqref{eq:36} suggests a correlation between the NST with the parameters $\beta^{\prime}$ and $K_{sym}(\rho_c)$. The correlation of the NST with $\beta^{\prime}$ and $K_{sym}(\rho_c)$ for the results of  finite range effective interaction are shown in Figures 14 and 15. As expected these quantities are highly correlated as depicted by the respective correlation coefficients. Linear fits to the correlations provide the relationships
\begin{eqnarray}
\varDelta r_{np} = (-0.09263\pm 0.00319) +(0.56413\pm 0.00637)\beta^{\prime}~ \text{fm}\label{eq:37},
\end{eqnarray}
and
\begin{equation}
\varDelta r_{np} =(0.27682\pm 0.0025)+(0.00085\pm 0.00002)K_{sym}(\rho_c)~ \text{fm}\label{eq:38}.
\end{equation}

\begin{figure}
\minipage{0.49\textwidth}
\centering
\includegraphics[width=85mm]{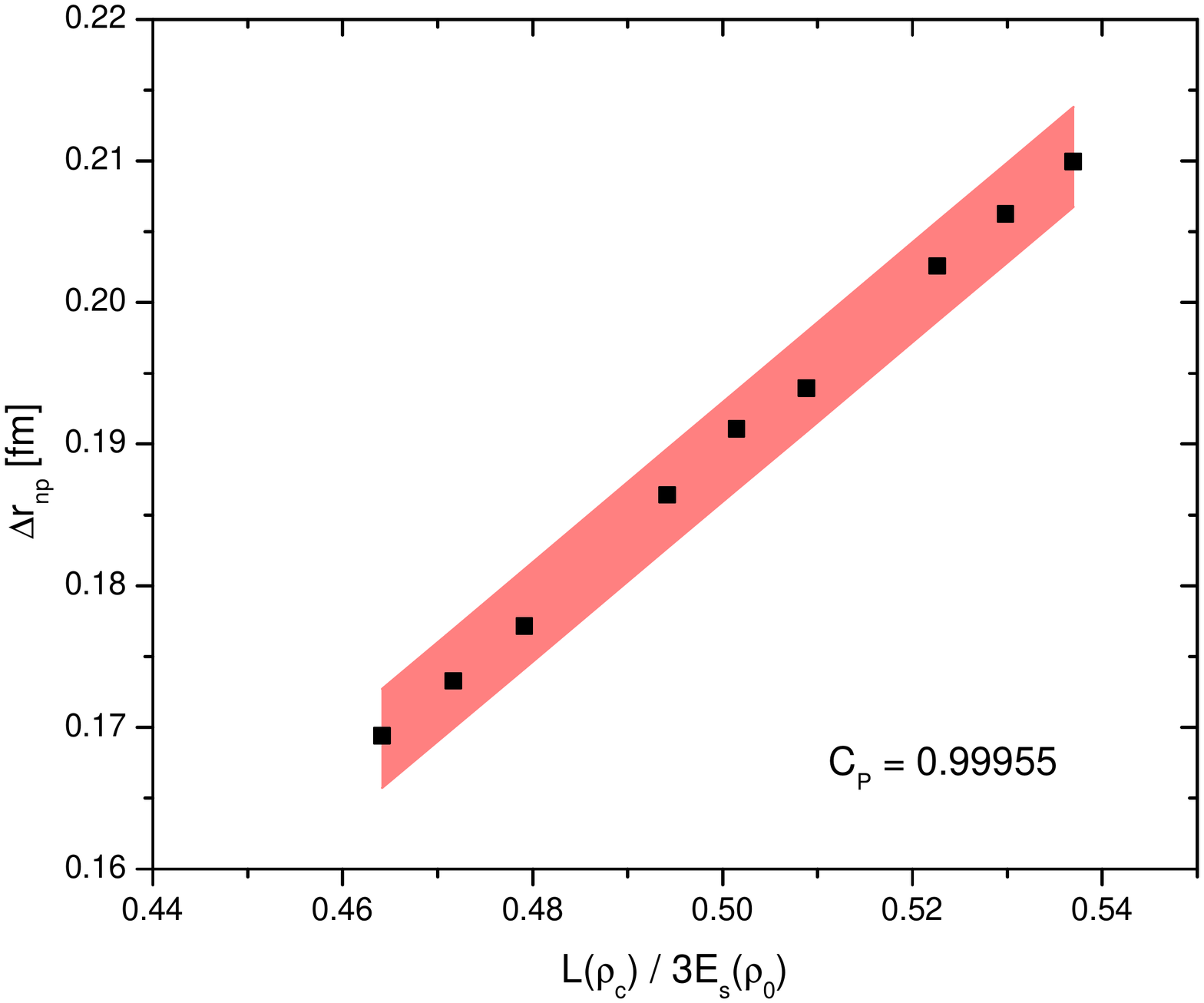}
\caption{Neutron skin thickness in $^{208}$Pb is correlated with the quantity $\beta^{\prime}=\frac{L(\rho_c)}{3E_s(\rho_0)}$. The shaded region shows the correlation band with a Pearson correlation coefficient $C_P=0.99955$.}
\endminipage
\hfill
\minipage{0.49\textwidth}
\includegraphics[width=85mm]{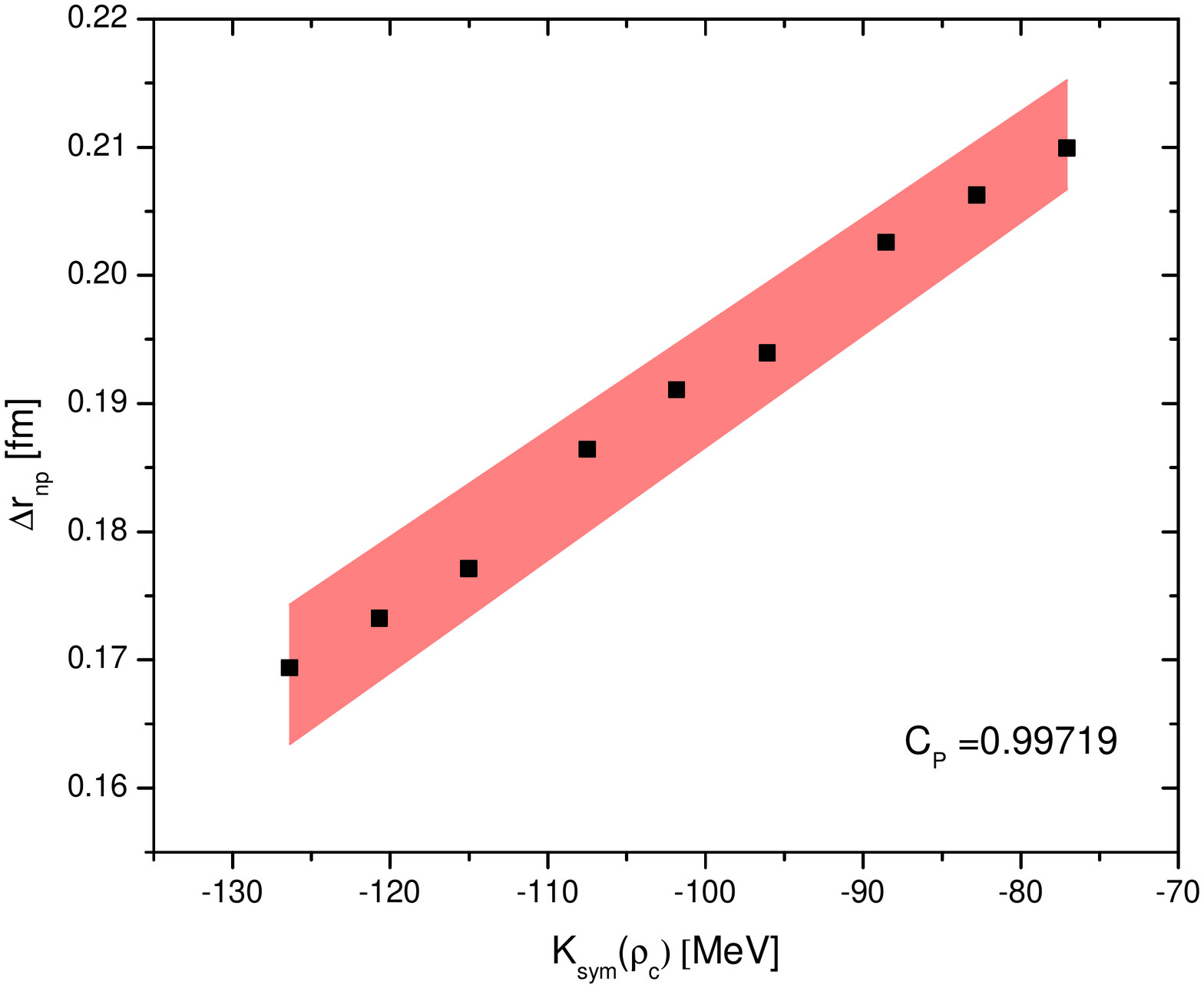}
\caption{Neutron skin thickness in $^{208}$Pb is correlated with the quantity $K_{sym}(\rho_c)$. The shaded region shows the correlation band with a Pearson correlation coefficient $C_P=0.99719$. }
\endminipage
\end{figure}

The neutron skin thickness $\varDelta r_{np}$ shows a strong correlation among certain isovector observables of heavy  finite nuclei such as $^{208}$Pb. Therefore accurate determination of the NST of $^{208}$Pb from experiments would provide constraints on the density dependence of NSE and on the slope parameter. The Lead Radius Experiment (PREX) \cite{prex2012}  is able to determine the neutron radius in $^{208}$Pb upto $1\%$ accuracy through the measurement of the parity violating asymmetry at low momentum transfer. The PREX results for the neutron skin thickness in $^{208}$Pb are  $\varDelta r_{np}=0.33^{+0.16}_{-0.18}$ fm \cite{prex2012}. Experiments with hadronic probes constrained the NST in $^{208}Pb$ as $\varDelta r_{np}= 0.16\pm (0.02)_{(\text{stat})}\pm (0.04)_{(\text{syst}}$ fm \cite{Klos2007} and  $\varDelta r_{np}= 0.211 ^{+0.054}_{-0.063}$ (Osaka-RCNP)\cite{Zenihiro2010}. Measurements from coherent pion photoproduction yield  a value $\varDelta r_{np}(^{208}Pb)= 0.15\pm 0.03$ fm  (Mainz experiment)\cite{Tarbert2014}. Our results from the finite range effective interactions are in conformity with these experimental results. In Figures 16 (a) and (b), we have shown the fitted results from our calculation and compared with those of the Mainz experiment \cite{Tarbert2014} and Osaka-RCNP results \cite{Zenihiro2010}. The Mainz results constrain the parameters $\beta$ and $\beta^{\prime}$ respectively in the range $0.145\leq \beta \leq 0.48$ and $0.38\leq \beta^{\prime} \leq 0.485$. Similarly, the constraints as obtained from a comparison of the Osaka-RCNP results with our results are $0.315\leq \beta \leq 0.93$ and $0.43\leq \beta^{\prime} \leq 0.625$. The quoted errors of $0.06$ fm for the NST as expected from PREX II are too large to make a contribution. Even the quoted errors of $0.03$ fm of the Mainz experiment define a $1\sigma$ interval of $0.06$ fm which spans the full range of cases plotted in Fig. 13. It is worth to mention here that, from an analysis of Skyrme type forces, Brown \cite{Brown2013} has obtained a linear relationship between the derivative of the NSE at $\rho_c$ and the neutron skin thickness of $^{208}$Pb as $\frac{dE_s(\rho)}{d\rho}|_{\rho_{c}}=p_{a}\varDelta r_{np}$, where $p_a=882\pm 32$ MeV fm$^2$. This relation can be transformed as $\beta^{\prime}=\frac{\rho_cp_a}{E_s(\rho_0)}\varDelta r_{np}$. For a given value of $E_s(\rho_0)=35$ MeV, we get $0.41\leq \beta^{\prime}\leq 0.735$ from the results of Osaka-RCNP and $0.33\leq \beta^{\prime}\leq 0.5$ from the results of Mainz experiment. These values are compatible with the constraints obtained from Fig.16.
\begin{figure}
\includegraphics[width=0.7\textwidth]{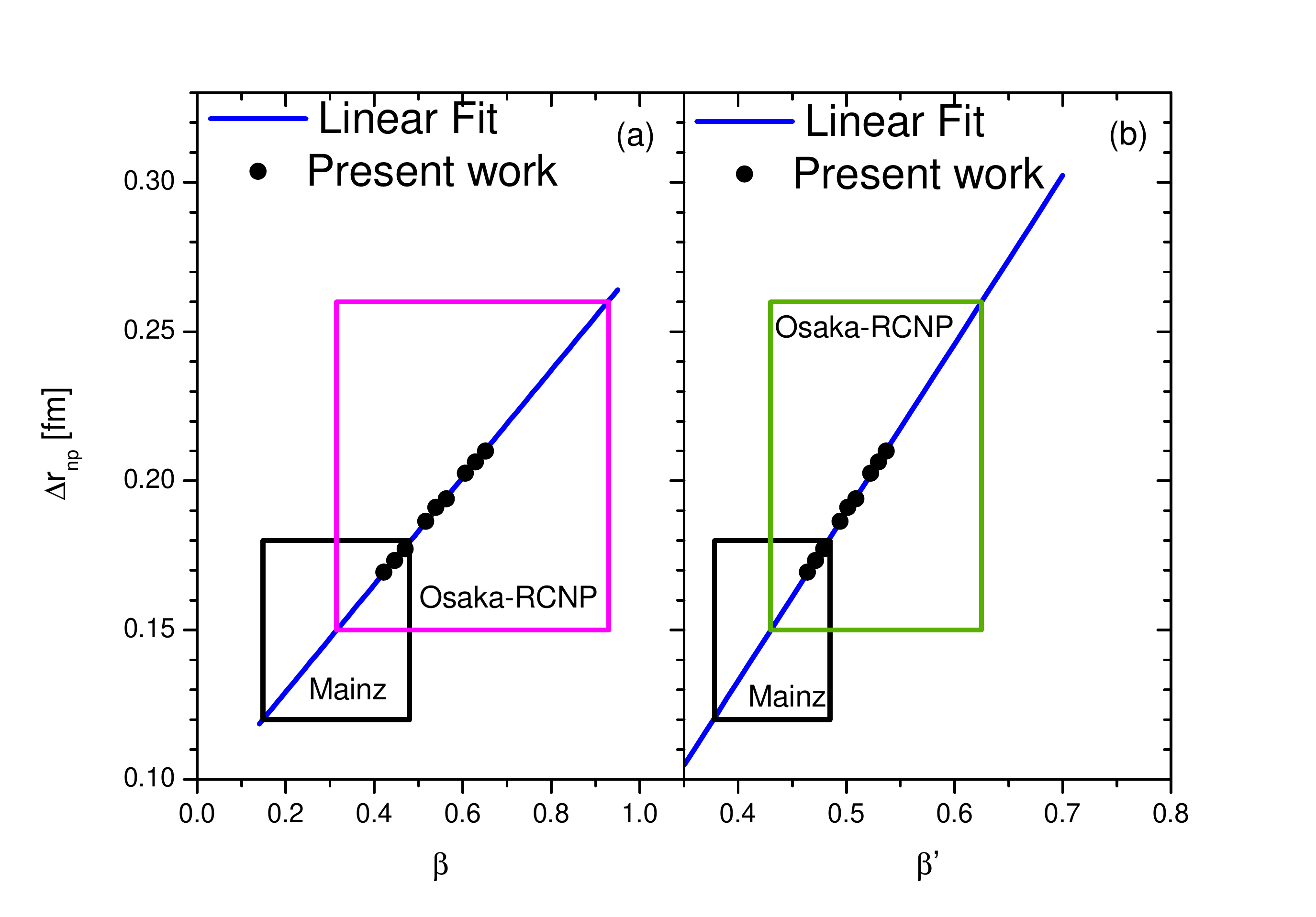}
\caption{(a)The neutron skin thickness of $^{208}$Pb is shown as a function of the parameter $\beta=\frac{L(\rho_0)}{3E_s(\rho_0)}$. (b) The same as a function of the parameter $\beta^{\prime}=\frac{L(\rho_c)}{3E_s(\rho_0)}$, ($\rho_c=0.11$ fm$^{-3}$). The solid dots are the calculations from the finite range effective interaction. The blue line is a linear fit to the solid dots. Our results are compared with the extracted results for NST of $^{208}$Pb from Mainz experiment \cite{Tarbert2014} and Osaka-RCNP  results\cite{Zenihiro2010}. }
\end{figure}

\section{Correlation of NST with Isovector Giant Dipole Resonance}
In medium-to-heavy neutron rich nuclei, the dipole response of a nucleus to an externally applied electric field is mostly dominated by the giant dipole resonance (GDR) of width $2-4$ MeV \cite{Berman1975, Piekarewicz2012}. The isovector giant dipole resonance is perceived as an out of phase collective oscillation of neutrons against protons. The excess neutrons in the neutron-rich nuclei may form a skin and oscillate against the isospin-saturated core giving rise to a low-energy $E1$  mode \cite{Paar2007}. The nuclear symmetry energy at subsaturation densities acts as a restoring force in this collective excitation phenomena. It is a well known fact that, the properties of IVGDR depend on the nuclear symmetry energy \cite{Krivine1984, Trippa2008, Lipparini1989, Piekarewicz2014, Tamii2011}.
Correlations with different collective excitation modes such as the IVGDR may also put some constraints on the density dependence of the nuclear symmetry energy. Correlations of nuclear symmetry energy parameters with other isovector modes of collective excitations such as electric dipole polarisability \cite{Maza2013, Maza2011, Birkhan2017, Piekarewicz2012, Zhang2014}, pygmy dipole resonance (PDR) \cite{Klimkiewicz2007, Baran2012, Daoutidis2011} and isovector giant quadrupole resonance (IVGQR) \cite{Maza2013a} also exist in literature.

The IVGDR energy constant can be expressed in a semiclassical framework as \cite{Blocki2013}
\begin{equation}
D=\sqrt{\frac{8\hbar^2}{mr_0^2}\left[\frac{E_s(\rho_0)}{1+3\frac{E_s(\rho_0)}{Q}A^{-1/3}}\right]}\label{eq:39}.
\end{equation}

The IVGDR energy constant depends on the factor $E_s(\rho_0)/Q$ and consequently depends on the neutron skin thickness $\varDelta r_{np}$ (see Eqs. \eqref{eq:29} and \eqref{eq:30}). Using the $Q$ values calculated in the present work, we obtained the IVGDR energy constant for $^{208}$Pb in the range $79.77-81.1$ MeV in close agreement with the experimental value $D_{exp}\approx 80$ MeV for heavy nuclei. In the left panel of Figure 17, we show $D$ as a function of the neutron skin thickness. The IVGDR energy constant appears to decrease with $\varDelta r_{np}$. Since the NST depends on the density slope parameter at a reference density, it is expected that, $D$ will have certain relation with $L(\rho_c)$. In the right panel of Figure 17, the IVGDR energy constant is plotted as a function of $L(\rho_c)$ obtained from the finite range effective interactions. It appears that $D$ decreases with an increase in the value of $L(\rho_c)$. The experimental value of $D$ is exactly reproduced for $L(\rho_c)=51.9$ MeV corresponding to a neutron skin thickness of $\varDelta r_{np}=0.194$ fm in $^{208}$Pb. 

We can write Eq. \eqref{eq:39} as
\begin{equation}
D=\sqrt{\frac{8\hbar^2}{mr_0^2}}\left(\frac{3}{A^{1/3}Q}\right)^{-1/2}\left[1+\frac{A^{1/3}Q}{3E_s(\rho_0)}\right]^{-1/2}.\label{eq:40}
\end{equation}

Expanding the square-bracketed term in powers of $\frac{A^{1/3}Q}{3E_s(\rho_0)}$ and retaining up to the 1st order we get
\begin{equation}
D\simeq \sqrt{\frac{8\hbar^2}{mr_0^2}}\left(\frac{3}{A^{1/3}Q}\right)^{-1/2}\left[1-\frac{A^{1/3}Q}{6E_s(\rho_0)}\right],\label{eq:41}
\end{equation}
which may be expressed as
\begin{equation}
D=B_A\left(\frac{a_{sym}(A)}{t}\right)^{1/2}\left[1-\frac{A^{1/3}Q}{6E_s(\rho_0)}\right].\label{eq:42}
\end{equation}
Here we have substituted the expression for $Q$ from the relationship of $t$ in Eq. \eqref{eq:30} and defined a constant  $B_A=\sqrt{\frac{4\hbar^2}{mr_0}A^{1/3}(I-I_c)}$ for  a given nucleus. In Eq.\eqref{eq:42}, the leading term is proportional to $\left(\frac{a_{sym}(A)}{t}\right)^{1/2}$. This expression in Eq.\eqref{eq:42} allows us to draw a linear correlation between the IVGDR energy and the quantity $\left(\frac{a_{sym}(A)}{t}\right)^{1/2}$. In Figure 18, we show $D$ as a function of $\left(\frac{a_{sym}(A)}{t}\right)^{1/2}$. The Pearson correlation coefficient for the present case is $C_P=0.82$. In the figure we show a linear fit to the correlated data points that reads as $D=73.833+0.538\left(\frac{a_{sym}(A)}{t}\right)^{1/2}$. Assuming $a_{sym}(A)=26.65$ MeV, with an IVGDR energy constant of $80$ MeV, we obtain $t=0.202$ fm. The linear fit formula provides a reasonable estimate of the NST for $^{208}$Pb.

\begin{figure}
\includegraphics[width=0.7\textwidth]{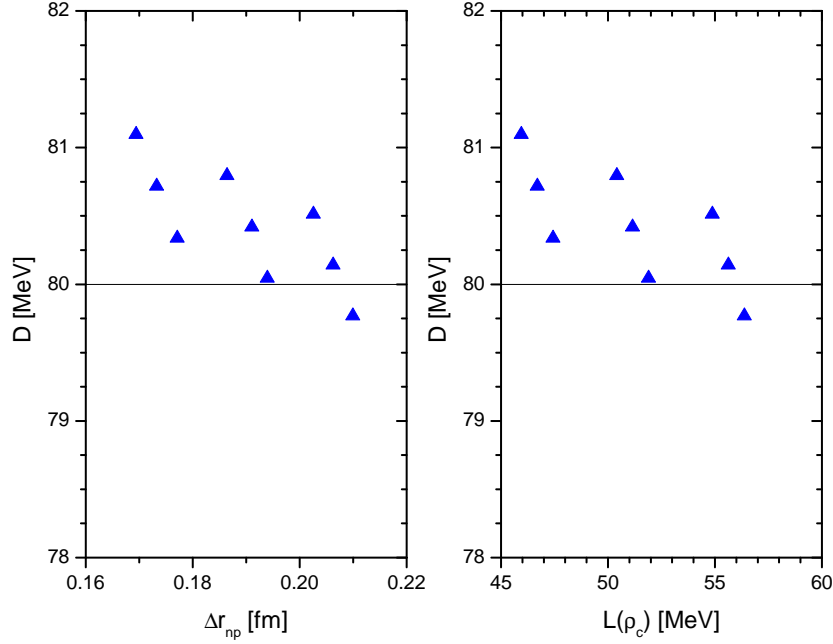}
\caption{(Left panel): The IVGDR energy constant $D$ in $^{208}$Pb is plotted as a function of neutron skin thickness. (Right panel): The IVGDR energy constant $D$ is shown as a function of the density slope parameter at a reference density $\rho_c=0.11$ fm $^{-3}$. The horizontal line shows the experimental value of $D$.}
\end{figure}
\begin{figure}
\includegraphics[width=0.7\textwidth]{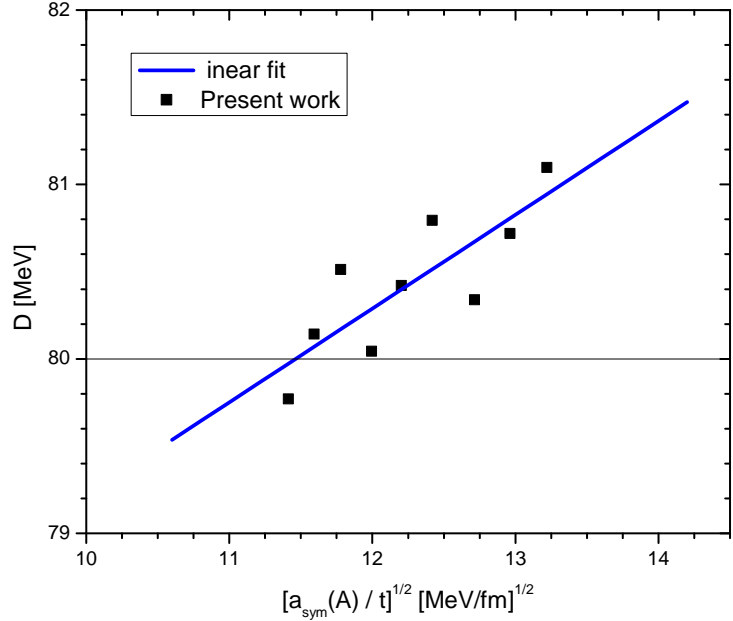}
\caption{The IVGDR energy is correlated with the quantity $\left(\frac{a_{sym}(A)}{t}\right)^{1/2}$. The horizontal line shows the experimental value of $D$. }
\end{figure}

\section{Conclusion}
In the present work, we have studied the density dependence of nuclear symmetry energy using a finite range effective interaction. The effective interaction has a finite range part so as to describe the correct momentum dependence of the nuclear mean field as extracted from the optical model fits in the heavy-ion collision studies at intermediate energies. By exploring the recently constrained nuclear symmetry energy at a subsaturation cross density ($\rho_c=0.11$ fm$^-3$), we have constructed three different sets of nuclear EoSs from the SEI. The EoSs constructed in the present work satisfactorily reproduce the properties of SNM at saturation density. We have considered the nuclear symmetry energy at saturation in the accepted range of $32\pm 4$ MeV and obtained the density slope parameter in the range $44\leq L(\rho_0)\leq 66$ MeV and the curvature parameter in the range $-201\leq K_{sym}(\rho_0) \leq -126$ MeV. These values are well within the constrained values from different analysis. The EoSs obtained in the present work not only satisfy the constraints for NSE in the subnormal densities but also pass within the experimentally extracted region of Russotto et al.\cite{Russotto2016}. The constraints coming from astrophysical and terrestrial laboratories limit the NSE at a density twice the normal nuclear matter density as $E_s(2\rho_0)= 46.9 \pm 10.1 $ MeV \cite{Zhang2019, Li2019}. The predictions for NSE from our EoSs lie well within these limits at $2\rho_0$ and therefore viable for different nuclear studies for a wide range of density domain. For the given range of $E_s(\rho_0)$ and $E_s(\rho_c)$, we have obtained  the density slope parameter in the range $46.69 \leq L(\rho_c) \leq 55.63$ MeV and the curvature symmetry energy at a subsaturation density in the range $-120.69\leq K_{sym}(\rho_c)\leq -82.80$ MeV. It is worth to mention here that while constraints on $L(\rho_c)$ are available in literature there are no reliable constraints on $K_{sym}(\rho_c)$. The predicted values of $L(\rho_c)$ are in conformity with the available constraints \cite{Zhang2014}.

We have analysed the correlation of the neutron skin thickness in $^{208}$Pb with different isovector parameters. In order to calculate the NST, we required the values of the symmetry coefficient $a_{sym}(A)$ as a function of mass number. We assumed the relationship $a_{sym}(A)\simeq E_s(\rho_A)$. It is worth to mention here that, this relation has already been verified by many workers for different mean field calculations. From the calculated values of $a_{sym}(A)$ we have extracted the values of surface symmetry energy coefficient in a leptodermous expansion. We were able to find a linear relationship between the surface symmetry energy coefficient with the density slope parameter at a subsaturation density $L(\rho_c)$. The NST is calculated within the framework of droplet model using the finite range effective interaction. Instead of correlating the NST with the usual density slope parameter at saturation or the quantity $E_s(\rho_0)-a_{sym}(A)$, we find it reasonable to have a correlation between $\varDelta r_{np}$ with the isovector indicators such as $1-\frac{a_{sym}(A)}{E_s(\rho_0)}$ and $\beta=\frac{E_s^{\prime}(\rho_0)}{E_s(\rho_0)}$. The NST as predicted from our EoSs lies in the range $0.17-0.21$ fm with a spread of $0.04 $ fm. These values are in conformity with the experimentally extracted values of NST from different works. We have explored the correlation between the NST  with the density slope parameter at a subsaturation density. A linear fit of $\varDelta r_{np}$ to the quantity $\beta^{\prime}=\frac{L(\rho_c)}{3E_s(\rho_0)}$ is obtained from the correlation procedure. A similar correlation between the NST and the curvature symmetry parameter at subsaturation density is obtained from the calculation.  From a comparison with the Mainz results, we constrained the parameters $\beta$ and $\beta^{\prime}$ respectively in the range $0.145\leq \beta \leq 0.48$ and $0.38\leq \beta^{\prime} \leq 0.485$. Similarly, the constraints as obtained from a comparison of the Osaka-RCNP results with our results are $0.315\leq \beta \leq 0.93$ and $0.43\leq \beta^{\prime} \leq 0.625$. We calculated the IVGDR energy constant in $^{208}$Pb and correlated the results with the NST and the density slope parameter at a subsaturation density. As a final remark, we say that, the parameters $\beta$ and $\beta^{\prime}$ should be accurately fixed up for an understanding of the NSE. Future experiments with accurate determination of the neutron skin thickness of nuclei may pin down these parameters.

%

\section*{Acknowledgement}
DB and SKT thank IUCAA, Pune (India) for providing hospitality and support during an academic visit where a part of this work is accomplished.


\begin{thebibliography}{99}

\section*{References}

\bibitem{Li2019a} B. A. Li, P. G. Krastev, D. H. Wen and N. B. Zhang, \textit{Eur. Phys. J. A}, \textbf{55}, 217 (2019).

\bibitem{Lattimer2000} J. Lattimer and M. Prakash, \textit{Phys. Rep.}, \textbf{333-334}, 121 (2000).

\bibitem{Steiner2005} A. W. Steiner, M. Prakash, J. Lattimer, P. Ellis, \textit{Phys. Rep.}, \textbf{41}, 325 (2005).

\bibitem{Steiner2012} A. W. Steiner, S. Gandolfi, \textit{Phys. Rev. Lett.}, \textbf{108}, 081102 (2012).

\bibitem{Ji2019} F. Ji, J. Hu, S. Bao and H. Shen, \textit{Phys. Rev. C}, \textbf{100}, 045801 (2019).

\bibitem{Lattimer2007} J. Lattimer and M. Prakash, \textit{Phys. Rep.}, \textbf{442}, 109 (2007).

\bibitem{Klahn2006} T. Kl\"{a}hn, D. Blaske, S. Typel, E. N. E. Van Dalen, A. Faessler, C. Fuchs, T. Gaitanos, H, Grigorian,  A. Ho, et al., \textit{Phys. Rev. C}, \textbf{74}, 035802 (2006).

\bibitem{Loan2011} D. T. Loan, N. H. Tan, D. T. Khoa, J. Margueron, \textit{Phys. Rev. C}, \textbf{83}, 065809 (2011).

\bibitem{Tsang2009} M. B. Tsang, Y. Zhang, P. Danielewicz, M. Famiano, Z. Li, W. G. Lynch and A. W. Steiner, \emph{Phys. Rev. Lett.}, \textbf{102}, 122701 (2009).

\bibitem{Tsang2012} M. B. Tsang \textit{et al.}, \emph{Phys. Rev. C} \textbf{86}, 015803 (2012).

\bibitem{Piek2012} J. Piekarewicz \textit{et al.}, \textit{Phys. Rev. C}, \textbf{85}, 041302(R) (2012).

\bibitem{Tamii2011} A. Tamii, I. Poltoratska, P. von Neumann-Cosel, Y. Fujita, T. Adachi, C. A. Bertulani, J. Carter, M. Dozono, H. Fujita, K. Fujita et al., \emph{Phys. Rev. Lett.}, \textbf{107}, 062502 (2011).

\bibitem{Maza2013} X. Roca-Maza, M. Brenna, G. Col\`{o}, M. Centelles, X. Vi\~{n}as, B. K. Agrawal, N. Paar, D. Vretenar and J. Piekarerwicz, \emph{Phys. Rev. C}, \textbf{88}, 024316 (2013).

\bibitem{Zhang2014} Z. Zhang and L. W. Chen, \emph{Phys. Rev. C}, \textbf{90}, 064317 (2014).

\bibitem{Gor2009} S. Goriely, N. Chamel, J M Pearson, \textit{Phys. Rev. Lett.}, \textbf{102}, 152503 (2009).

\bibitem{Wang2014} N. Wang, M. Liu, X. Z. Wu, J. Meng, \textit{Phys. Lett. B}, \textbf{734}, 215 (2014).

\bibitem{Tian2014} J. L. Tian, H. T. Cui, K . K. Zheng, N. Wang \textit{Phys. Rev. C},\textbf{90}, 024313 (2014).

\bibitem{Chen2015} W. C. Chen, J. Piekarewicz, \textit{Phys. Lett. B}, \textbf{748}, 284 (2015).

\bibitem{Warda2009} M. Warda, X. Vi\~{n}as, X. Roca-Maza and M. Centelles, \textit{Phys. Rev. C}, \textbf{80}, 024316 (2009).

\bibitem{Vinas2014} X. Vi\~{n}as, M. Centelles, X. Roca-Maza, and M. Warda, \emph{Eur. Phys. J. A}, \textbf{50}, 27 (2014).

\bibitem{Mondal2016} C. Mondal, B. K. Agrawal, M. Centelles, G. Col\`{o}, X. Roca-Maza, N. Paar, X. Vi\~{n}as, S. K. Singh and S. K. Patra, \emph{Phys. Rev. C}, \textbf{93}, 064303 (2016).

\bibitem{Wang2013}N. Wang, L. Ou, M. Liu, \textit{Phys. Rev. C}, \textbf{87}, 034327 (2013).

\bibitem{Zhang2013} Z. Zhang and L. W. Chen, \emph{Phys. Lett. B}, \textbf{726}, 234 (2013).


\bibitem{BKA12}B. K. Agrawal, J. N. De and S. K. Samaddar, \textit{Phys. Rev. Lett.}, \textbf{109}, 262501 (2012).

\bibitem{BKA13}B. K. Agrawal, J. N. De, S. K. Samaddar, G. Col\`{o}, A. Sulaksono, \textit{Phys. Rev. C}, \textbf{87}, 051306(R) (2013).

\bibitem{Hashimoto2015} T. Hashimoto \textit{et al.}, \textit{Phys. Rev. C}, \textbf{92}, 031305 (2015).

\bibitem{Tonchev2017} A. Tonchev, N. Tsoneva, C. Bhatia, A. Arnold, S. Goriely, \textit{et al.} , \textit{Phys. Lett. B}, \textbf{773}, 20 (2017).

\bibitem{Tamii2014} A. Tamii, P. von Neumann Cosel and I. Poltoratska, \textit{Eur. Phys. J. A}, \textbf{50}, 28 (2014).

\bibitem{Li2013} B. A. Li and X. Han, \textit{Phys. Lett. B}, \textbf{727}, 276 (2013).

\bibitem{Oertel2017}M. Oertel, M. Hempel, T. Kl\"{a}hn, S. Typel, \textit{Rev. Mod. Phys.}, \textbf{89}, 015007 (2017).

%

\bibitem{Trzcinska2001} A. Trzci\'{n}ska, J. Jastrz\k{e}bski, P. Lubi\'{n}ski, F. J. Hartmann, R. Schmidt, T. von Egidy, B. Kos, \textit{Phys. Rev. Lett.}, \textbf{87}, 082501 (2001).

\bibitem{Brown2007} B. A. Brown, G. Shen, G. C. Hillhouse, J. Meng and A. Trzci\'{n}ska, \textit{Phys. Rev. C}, \textbf{87}, 034305 (2007).

\bibitem{Klos2007} B. Klos \textit{et al.}, \emph{Phys. Rev. C}, \textbf{76}, 014311 (2007).

\bibitem{Zenihiro2010} J. Zenihiro, \textit{et al.},\emph{Phys. Rev. C}, \textbf{82}, 044611 (2010). 

\bibitem{Roca-Maza2012} X. Roca-Maza, B. K. Agrawal, G. Col\`{o}, W. Nazarewicz, N. Paar, J. Piekarewicz, P. G. Reinhard and D. Vretenar, \textit{AIP Conf. Proc.}, \textbf{1491}, 204 (2012).

\bibitem{prex2012} S. Abrahamyan, Z. Ahmed et al. (PREX collaboration), \emph{Phys. Rev. Lett.}, \textbf{108}, 112502 (2012).

\bibitem{prex2} K. Paschke \textit{et al.}, Jefferson Lab Experiment E12-11-101 (PREX-II) proposal at http://hallaweb.jlab.org/parity/prex (2014).

\bibitem{Myers1980} W. D. Myers and W. J. \'{S}wia\c{t}ecki, \textit{Nucl. Phys. A}, \textbf{336}, 267 (1980).

\bibitem{Centelles2010} M. Centelles, X. Roca-Maza, X. Vi\~{n}as and M. Warda, \emph{Phys. Rev. C}, \textbf{82}, 054314 (2010).

\bibitem{Maza2011} X. Roca-Maza, M. Centelles, X. Vi\~{n}as, M. Warda, \emph{Phys. Rev. Lett.}, \textbf{106}, 252501 (2011).

\bibitem{Zhou2019} Y. Zhou and L. W. Chen, \emph{Astrophys. J.}, \textbf{886}, 52 (2019).
%
%
%
%
%

\bibitem{Behera2016} B. Behera, T. R. Routray and S. K. Tripathy, \textit{Mod. Phys. Lett. A}, \textbf{31}, 1650194 (2016).

\bibitem{Berstch1988} G. F. Bertsch, S. Das Gupta, \emph{Phys. Rep. }, \textbf{160}, 189 (1988).

\bibitem{Kuper1974} W. A. Kuper, G. Wegman and E. R. Hilf, \emph{ Ann. Phys.(NY)}, \textbf{88}, 454 (1974).

\bibitem {Pan1993} Q. Pan, P. Danielewicz, \textit{Phys. Rev. Lett.}, \textbf{70}, 2062 (1993).

\bibitem{Mota1992} V. de La Mota, F. S\'{e}bille, B. Remand, P. Schuck, \textit{Phys. Rev. C}, \textbf{46}, 667 (1992).

\bibitem{Zhang1994} J. Zhang, S. Das Gupta, C. Gale, \textit{Phys. Rev. C} \textbf{50}, 1617 (1994).

\bibitem{Haddad1995} F. Haddad, F. S\'{e}bille, M. Marine, V. de La Mota, P. Schuck, B. Johault, \textit{Phys. Rev. C}, \textbf{52}, 2013 (1995).

\bibitem{Dan1998} P. Danielewicz, \emph{Phys. Rev. Lett.}, \textbf{81}, 2438 (1998).

\bibitem{Dan2000} P. Danielewicz, \emph{Nucl. Phys. A}, \textbf{673}, 375 (2000).


\bibitem{Routray2000} T. R. Routray, B. Sahoo, R. K. Satpathy and B. Behera, \emph{J. Phys. G: Nucl.Part. Phys.}, \textbf{26}, 887 (2000).

\bibitem{Behera1998} B. Behera, T. R. Routray and  R. K. Satpathy, \emph{J. Phys. G: Nucl.Part. Phys.}, \textbf{24}, 2073 (1998).

\bibitem{Behera2002} B. Behera, T. R. Routray, B. Sahoo and  R. K. Satpathy, \emph{Nucl. Phys. A}, \textbf{699}, 770 (2002).

\bibitem{Behera2005} B. Behera, T. R. Routray and  A. Pradhan, \emph{Mod. Phys. Lett. A}, \textbf{20}, 2639 (2005).

\bibitem{Behera2009} B. Behera, T. R. Routray and  S. K. Tripathy, \emph{J. Phys. G: Nucl.Part. Phys.}, \textbf{36}, 125105 (2009).

\bibitem{Behera2011} B. Behera, T. R. Routray and  S. K. Tripathy, \emph{J. Phys. G: Nucl.Part. Phys.}, \textbf{38}, 115104 (2011).

\bibitem{Routray2011} T. R. Routray, S. K. Tripathy, B. B Dash, B. Behera and D. N. Basu, \emph{Eur. Phys. J. A}, \textbf{47}, 92 (2011).

\bibitem{Behera2007} B. Behera, T. R. Routray, A. Pradhan, S. K. Patra and P. K. Sahu, \emph{Nucl. Phys. A}, \textbf{794}, 132 (2007).

\bibitem{Baran2005} V. Baran, M. Colonna, V. Greco and M. Di Toro, \emph{Phys. Rep. }, \textbf{410}, 335 (2005).


\bibitem{Li2008} B. A. Li, L. W. Chen and C. M. Ko, \emph{Phys. Rep. }, \textbf{464}, 113 (2008).


\bibitem{Lattimer2012} J. Lattimer, \emph{Annu. Rev. Nucl. Part. Sci. }, \textbf{62}, 485 (2012).

\bibitem{Horowicz2014} C. J. Horowitz \textit{et al.}, \emph{J. Phys. G: Nucl.Part. Phys.}, \textbf{41}, 093001 (2014).

\bibitem{Dutra2012} M. Dutra, O. Lourenco, J. S. S. Martins, A. Delfino, J. R. Stone and P. D. Stevenson, \emph{Phys. Rev. C}, \textbf{85}, 035201 (2012).



%
%
%

\bibitem{Wang2015a} N. Wang, M. Liu, L. Ou and Y. Zhang, \emph{Phys. Lett. B}, \textbf{751}, 553 (2015).

\bibitem{Moller2012} P. M\"{o}ller, W. D. Myers, H. Sagawa and S. Yoshida, \emph{Phys. Rev.  Lett.}, \textbf{108}, 052501(2012).

\bibitem{Lattimer2013} J. Lattimer and Y. Lim, \emph{Astrophys. J.}, \textbf{771}, 51 (2013).

\bibitem{Dan2009} P. Danielewicz and J. Lee, \emph{Nucl. Phys. A}, \textbf{818}, 36 (2009).


\bibitem{Maza2015} X. Roca-Maza, X. Vi\~{n}as, M. Centelles, B. K. Agrawal, G. Col\`{o}, N. Paar, J. Piekarewicz and D. Vretenar, \emph{Phys. Rev. C}, \textbf{92}, 064304 (2015).


\bibitem{Rossi2013} D. M. Rossi, P. Adrich, F. Aksouh, H. Alvarez-Pol, T. Aumann, J. Benlliure, M. Bohmer, K. Boretzky, E. Casarejos, M. Chartier et al., \emph{Phys. Rev. Lett.}, \textbf{111}, 242503 (2013).

\bibitem{Chen2010} L. W. Chen, C. M. Ko, B. A. Li and J. Xu, \emph{Phys. Rev. C}, \textbf{82}, 024321 (2010).

\bibitem{Fuchs2006} C. Fuchs and H. H. Wolter, \textit{Eur. Phys. J. A}, \textbf{30}, 5 (2006).

\bibitem{Brown2013} B. A. Brown, \emph{Phys. Rev. Lett.}, \textbf{111}, 232502 (2013).

\bibitem{Abott2018} B. P. Abott \textit{et al.} (LIGO and VIRGO collaborations), \emph{Phys. Rev. Lett.}, \textbf{121}, 161101 (2018).

\bibitem{Zhang2019} N. B. Zhang and B. A. Li, \emph{Eur. Phys. J. A}, \textbf{55}, 39 (2019).

\bibitem{Li2019} B. A. Li, P. G. Krastev, D. H. Wen, W. J. Xie and N. B. Zhang, \emph{AIP conference proceedings}, \textbf{2127}, 020018 (2019).

\bibitem{Tong2020} H. Tong, P. Zhao and J. Meng, \emph{Phys. Rev. C}, \textbf{101}, 035802 (2020).

\bibitem{Xie2019} W. J. Xie and B. A. Li, \textit{Astrophys. J.}, \textbf{883}, 174 (2019).arXiv: 1907.10741 

\bibitem{Zhang2020} Y. Zhang, M. Liu, C. J. Xia, Z. Li and S. K. Biswal, \emph{Phys. Rev. C}, \textbf{101}, 034303 (2020).

%

\bibitem{nicer1} T. E. Riley, A. L. Watts, S. Bogdanov, \textit{et al.}, \emph{Astrophys. J. Lett.}, \textit{887}, L21 (2019).

\bibitem{nicer2} G. Raaijmakers, T. E. Riley, A. L. Watts, \textit{et al.}, \emph{Astrophys. J. Lett.}, \textit{887}, L22 (2019).

\bibitem{nicer3} M. C. Miller, F. K. Lamb, A. J. Dittmann, \textit{et al.}, \emph{Astrophys. J. Lett.}, \textit{887}, L24 (2019).

\bibitem{nicer4} S. Bogdanov, S. Guillot, P. S. Ray, \textit{et al.}, \emph{Astrophys. J. Lett.}, \textit{887}, L25 (2019).

\bibitem{nicer5} S. Bogdanov, F. K. Lamb, S. Mahmoodifar, \textit{et al.}, \emph{Astrophys. J. Lett.}, \textit{887}, L26 (2019).

\bibitem{Cromartie2019} H. T. Cromartie, E. Fonseca, S. M. Ransom, \textit{et al.}, \emph{NatAs}, \textbf{2}, 72 (2020).

\bibitem{Russotto2016} P. Russotto, S. Gannon \textit{et al.}, \emph{Phys. Rev. C}, \textbf{94}, 034608 (2016).

\bibitem{Russotto2011} P. Russotto  \textit{et al.}, \emph{Phys. Lett. B}, \textbf{697}, 471 (2011).

\bibitem{Cozma2013}M. D. Cozma, Y. Leifels, W. Trautmann, Q. Li and  P. Russotto, \textit{Phys. Rev. C}, \textbf{88}, 044912 (2013).


\bibitem{Lattimer2014} J. M. Lattimer and A. W. Steiner, \textit{Eur. Phys. J. A}, \textbf{50}, 40 (2014).


\bibitem{Centelles2009} M. Centelles, X. Roca-Maza, X. Vi\~{n}as and M. Warda, \textit{Phys. Rev. Lett.}, \textbf{102}, 122502 (2009).

\bibitem{Brack1985} M. Brack, C. Guet and H. B. Hakansson, \textit{Phys. Rep.}, \textbf{123}, 275 (1985).

\bibitem{Centelles1998} M. Centelles, M. Del Estal and X. Vi\~{n}as, \textit{Nucl. Phys. A}, \textbf{635}, 193 (1998).

\bibitem{Treiner1986} J. Treiner and H. Krivine, \textit{Ann. Phys.}, \textbf{170}, 406 (1986).


\bibitem{Chen2011} L. W. Chen, \textit{Phys. Rev. C}, \textbf{83}, 044308 (2011). arXiv:1101.5217

\bibitem{Chabanat1998} E. Chabanat, P.Bonche, P. Haensel, J. Meyer and R. Schaeffer, \textit{Nucl. Phys. A}, \textbf{635}, 231 (1998).

\bibitem{Wang2015} N. Wang, M. Liu, H. Jiang, J. L. Tian and Y. M. Zhao, \emph{Phys. Rev. C}, \textbf{91}, 044308 (2015).

\bibitem{Raduta2018} Ad. R. Raduta and F. Gulminelli, \textit{Phys. Rev. C}, \textbf{97}, 064309 (2018).arXiv: 1712.05973

\bibitem{Zhang2018} Z. Zhang, Y. Lim, J. W. Holt and C. M. Ko, \textit{Phys. Lett. B}, \textbf{777}, 73 (2018).

\bibitem{Tarbert2014} C. M. Tarbert \textit{et al.} ( Crystal Ball at MAMI and A2 Collaboration), \textit{Phys. Rev. Lett.}, \textbf{112}, 242502 (2014).

\bibitem{Berman1975} B. L. Berman and S. C. Fultz, \textit{Rev. Mod. Phys.}, \textbf{47}, 413 (1975).

\bibitem{Piekarewicz2012} J. Piekarewicz, B. K. Agrawal, G. Col\`{o}, W. Nazarewicz, N. Paar, P. -G. Reinhard, X. Roca-Maza and D. Vretenar, \textit{Phys. Rev. C}, \textbf{85}, 041302(R) (2012).

\bibitem{Paar2007} N. Paar \textit{et al.}, \textit{Rep. Prog. Phys.}, \textbf{70}, 691 (2007).

\bibitem{Krivine1984} H. Krivine, J. Treiner, O. Bohigas, \textit{Nucl. Phys. A}, \textbf{336 }, 155 (1984).

\bibitem{Trippa2008}L. Trippa, G. Col\`{o}, E. Vigezzi, \emph{Phys. Rev. C}, \textbf{77}, 061304(R) (2008).

\bibitem{Lipparini1989}E. Lipparini, S. Stringari, \emph{Phys. Rep.}, \textbf{175}, 103 (1989).

\bibitem{Piekarewicz2014} J. Piekarewicz, \textit{Eur. Phys. J. A}, \textbf{50}, 25 (2014).

\bibitem{Birkhan2017} J. Birkhan, M. Miorelli, S. Bacca, S. Bassauer, \textit{et al.}, \textit{Phys. Rev. Lett.}, \textbf{118}, 252501 (2017).

\bibitem{Klimkiewicz2007} A. Klimkiewicz, N. Paar, P. Adrich, M. Fallot, \textit{et al.}, \textit{Phys. Rev. C}, \textbf{76}, 051603 (2007).


\bibitem{Daoutidis2011} I. Daoutidis and S. Goriely, \textit{Phys. Rev. C}, \textbf{84}, 027301 (2011).

\bibitem{Baran2012} V. Baran, B. Frecus, M. Colonna, and M. Di Toro, \textit{Phys. Rev. C}, \textbf{85}, 051601(R) (2012).

\bibitem{Maza2013a} X. Roca-Maza, M. Brenna, B. K. Agrawal, P. F. Bortignon, G. Col\`{o}, Li-Gang Cao, N. Paar  and D. Vretenar, \emph{Phys. Rev. C}, \textbf{87}, 034301 (2013).


\bibitem{Blocki2013} J. P. Blocki, A. G. Magner, P. Ring and A. A. Vlasenko, \emph{Phys. Rev. C}, \textbf{87}, 044304 (2013).


\end{thebibliography}
\end{document}